\documentclass[12pt]{article}
\usepackage{epsfig}
\usepackage{amsmath}
\usepackage{hhline}
\usepackage{amssymb}
\usepackage{times}
\usepackage{cite}
\usepackage{multirow}

\newlength{\dinwidth}
\newlength{\dinmargin}
\begin{document}  
% The rest
\newcommand{\pom}{{I\!\!P}}
\newcommand{\reg}{{I\!\!R}}
\newcommand{\slowpi}{\pi_{\mathit{slow}}}
\newcommand{\fiidiii}{F_2^{D(3)}}
\newcommand{\fiidiiiarg}{\fiidiii\,(\beta,\,Q^2,\,x)}
\newcommand{\n}{1.19\pm 0.06 (stat.) \pm0.07 (syst.)}
\newcommand{\nz}{1.30\pm 0.08 (stat.)^{+0.08}_{-0.14} (syst.)}
\newcommand{\fiidiiiful}{F_2^{D(4)}\,(\beta,\,Q^2,\,x,\,t)}
\newcommand{\fiipom}{\tilde F_2^D}
\newcommand{\ALPHA}{1.10\pm0.03 (stat.) \pm0.04 (syst.)}
\newcommand{\ALPHAZ}{1.15\pm0.04 (stat.)^{+0.04}_{-0.07} (syst.)}
\newcommand{\fiipomarg}{\fiipom\,(\beta,\,Q^2)}
\newcommand{\pomflux}{f_{\pom / p}}
\newcommand{\nxpom}{1.19\pm 0.06 (stat.) \pm0.07 (syst.)}
\newcommand {\gapprox}
   {\raisebox{-0.7ex}{$\stackrel {\textstyle>}{\sim}$}}
\newcommand {\lapprox}
   {\raisebox{-0.7ex}{$\stackrel {\textstyle<}{\sim}$}}
\def\gsim{\,\lower.25ex\hbox{$\scriptstyle\sim$}\kern-1.30ex%
\raise 0.55ex\hbox{$\scriptstyle >$}\,}
\def\lsim{\,\lower.25ex\hbox{$\scriptstyle\sim$}\kern-1.30ex%
\raise 0.55ex\hbox{$\scriptstyle <$}\,}
\newcommand{\pomfluxarg}{f_{\pom / p}\,(x_\pom)}
\newcommand{\dsf}{\mbox{$F_2^{D(3)}$}}
\newcommand{\dsfva}{\mbox{$F_2^{D(3)}(\beta,Q^2,x_{I\!\!P})$}}
\newcommand{\dsfvb}{\mbox{$F_2^{D(3)}(\beta,Q^2,x)$}}
\newcommand{\dsfpom}{$F_2^{I\!\!P}$}
\newcommand{\gap}{\stackrel{>}{\sim}}
\newcommand{\lap}{\stackrel{<}{\sim}}
\newcommand{\fem}{$F_2^{em}$}
\newcommand{\tsnmp}{$\tilde{\sigma}_{NC}(e^{\mp})$}
\newcommand{\tsnm}{$\tilde{\sigma}_{NC}(e^-)$}
\newcommand{\tsnp}{$\tilde{\sigma}_{NC}(e^+)$}
\newcommand{\st}{$\star$}
\newcommand{\sst}{$\star \star$}
\newcommand{\ssst}{$\star \star \star$}
\newcommand{\sssst}{$\star \star \star \star$}
\newcommand{\tw}{\theta_W}
\newcommand{\sw}{\sin{\theta_W}}
\newcommand{\cw}{\cos{\theta_W}}
\newcommand{\sww}{\sin^2{\theta_W}}
\newcommand{\cww}{\cos^2{\theta_W}}
\newcommand{\trm}{m_{\perp}}
\newcommand{\trp}{p_{\perp}}
\newcommand{\trmm}{m_{\perp}^2}
\newcommand{\trpp}{p_{\perp}^2}
\newcommand{\alp}{\alpha_s}

\newcommand{\alps}{\alpha_s}
\newcommand{\sqrts}{$\sqrt{s}$}
\newcommand{\LO}{$O(\alpha_s^0)$}
\newcommand{\Oa}{$O(\alpha_s)$}
\newcommand{\Oaa}{$O(\alpha_s^2)$}
\newcommand{\PT}{p_{\perp}}
\newcommand{\JPSI}{J/\psi}
\newcommand{\sh}{\hat{s}}
\newcommand{\uh}{\hat{u}}
\newcommand{\MP}{m_{J/\psi}}
\newcommand{\PO}{I\!\!P}
\newcommand{\xbj}{x}
\newcommand{\xpom}{x_{\PO}}
\newcommand{\ttbs}{\char'134}
\newcommand{\xpomlo}{3\times10^{-4}}  
\newcommand{\xpomup}{0.05}  
\newcommand{\dgr}{^\circ}
\newcommand{\pbarnt}{\,\mbox{{\rm pb$^{-1}$}}}
\newcommand{\gev}{\,\mbox{GeV}}
\newcommand{\WBoson}{\mbox{$W$}}
\newcommand{\fbarn}{\,\mbox{{\rm fb}}}
\newcommand{\fbarnt}{\,\mbox{{\rm fb$^{-1}$}}}
%
% Some useful tex commands
%
\newcommand{\qsq}{\ensuremath{Q^2} }
\newcommand{\gevsq}{\ensuremath{\mathrm{GeV}^2} }
\newcommand{\et}{\ensuremath{E_t^*} }
\newcommand{\rap}{\ensuremath{\eta^*} }
\newcommand{\gp}{\ensuremath{\gamma^*}p }
\newcommand{\dsiget}{\ensuremath{{\rm d}\sigma_{ep}/{\rm d}E_t^*} }
\newcommand{\dsigrap}{\ensuremath{{\rm d}\sigma_{ep}/{\rm d}\eta^*} }
% Journal macro
\def\Journal#1#2#3#4{{#1} {\bf #2} (#3) #4}
\def\NCA{\em Nuovo Cimento}
\def\NIM{\em Nucl. Instrum. Methods}
\def\NIMA{{\em Nucl. Instrum. Methods} {\bf A}}
\def\NPB{{\em Nucl. Phys.}   {\bf B}}
\def\PLB{{\em Phys. Lett.}   {\bf B}}
\def\PRL{\em Phys. Rev. Lett.}
\def\PRD{{\em Phys. Rev.}    {\bf D}}
\def\ZPC{{\em Z. Phys.}      {\bf C}}
\def\EJC{{\em Eur. Phys. J.} {\bf C}}
\def\CPC{\em Comp. Phys. Commun.}
%
% Some font size definitions for this paper
%
\newcommand{\tablesize}{\small}
\newcommand{\captiontext}{\small \it}

\begin{titlepage}

\noindent
\begin{flushleft}
DESY 06-029\hfill ISSN 0418-9833\\
April 2006
\end{flushleft}

\vspace*{2cm}

\begin{center}
\begin{Large}

{\bf Tau Lepton Production in \boldmath{$ep$} Collisions at HERA}

\vspace{2cm}

H1 Collaboration

\end{Large}
\end{center}

\vspace{2cm}

\begin{abstract}
\noindent
The production of tau leptons in $ep$ collisions is investigated using data recorded by the H1 detector at HERA in the period 1994-2000.
Tau leptons are identified by detecting their decay products, using leptonic and hadronic decay modes. 
The cross section for the production of tau lepton pairs is measured for the first time at HERA. Furthermore, a search for events with an energetic  isolated tau lepton and with large missing transverse momentum is performed.
The results are found to be in agreement with the Standard Model predictions. 
\end{abstract}

\vspace{1.5cm}
\begin{center}
Submitted to Eur.~Phys.~J.~C
\end{center}

\end{titlepage}

%%% all tex files here
\noindent
%-- H1AUTS Author list by names 
%-- Status: Mon Mar  6 16:07:34 CET 2006  Number of authors = 295 
A.~Aktas$^{9}$,                %DESY-PD        09/05           Aktas               
V.~Andreev$^{25}$,             %LPI -PD        8/88            Andreev             
T.~Anthonis$^{3}$,             %ANTW-ST        11/99           Anthonis            
B.~Antunovic$^{26}$,           %MPIM-ST        09/03           Antunovic           
S.~Aplin$^{9}$,                %DESY-PD        01/04           Aplin               
A.~Asmone$^{33}$,              \linebreak %ROME-ST        07/2            Asmone              
A.~Astvatsatourov$^{3}$,       %BRUX-PD        07/04           Astvatsatourov      
A.~Babaev$^{24, \dagger}$,     %ITEP-LEFT      12/05           Babaev              
S.~Backovic$^{30}$,            %PODG-PD        03/2            Backovic            
A.~Baghdasaryan$^{37}$
% ...

%-- H1 Institutes 
\bigskip{\it \noindent
% ...
$ ^{3}$ Inter-University Institute for High Energies ULB-VUB, Brussels;
          Universiteit Antwerpen, Antwerpen; Belgium$^{ c}$ \\
% ...
$ ^{9}$ DESY, Hamburg, Germany \\
% ...
$ ^{24}$ Institute for Theoretical and Experimental Physics,
          Moscow, Russia$^{ k}$ \\
 $ ^{25}$ Lebedev Physical Institute, Moscow, Russia$^{ e}$ \\
 $ ^{26}$ Max-Planck-Institut f\"ur Physik, M\"unchen, Germany \\
% ...
$ ^{30}$ Faculty of Science, University of Montenegro,
          Podgorica, Serbia and Montenegro$^{ e}$ \\
% ...
$ ^{33}$ Dipartimento di Fisica Universit\`a di Roma Tre
          and INFN Roma~3, Roma, Italy \\
% ...
$ ^{37}$ Yerevan Physics Institute, Yerevan, Armenia \\
% ...
$ ^{\dagger}$ Deceased \\

\bigskip { \noindent
% ...
$ ^c$ Supported by FNRS-FWO-Vlaanderen, IISN-IIKW and IWT
      and  by Interuniversity
Attraction Poles Programme,
      Belgian Science Policy \\
% ...
$ ^e$ Supported by the Deutsche Forschungsgemeinschaft \\
% ...
$ ^k$ Partially Supported by Russian Foundation
      for Basic Research,  grants  03-02-17291
      and  04-02-16445 \\
}
}
\newpage
%======================================
\section{Introduction} 
%======================================

\begin{figure}[h]
\setlength{\unitlength}{1mm}
\begin{center}
\epsfig{file=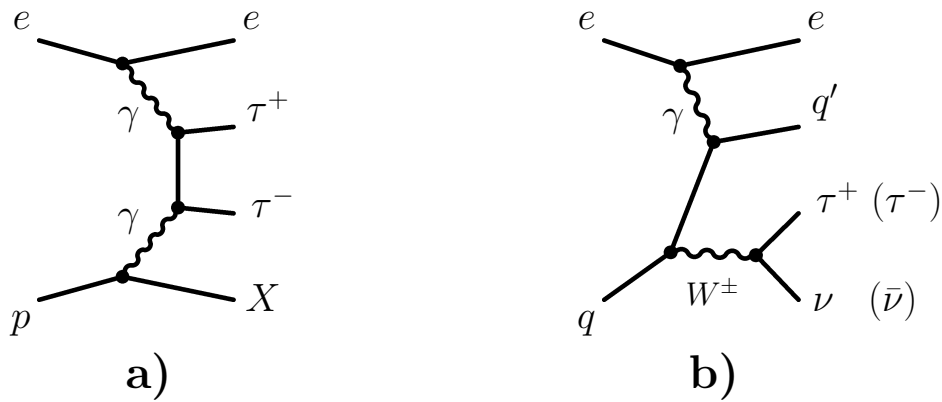,width=13.0cm,clip=}
\caption{Diagrams of the main production mechanisms of tau leptons in electron--proton collisions: {\bf a)} tau pair production via photon--photon collisions and {\bf b}) single $W$ boson production followed by the subsequent decay of the $W$ into a tau  lepton and a tau anti--neutrino. }
\label{diag}
\end{center}
\end{figure}

In the Standard Model (SM), tau leptons are  produced 
either in pairs or in association with a tau anti--neutrino, as expected from lepton flavour conservation. In electron--proton collisions,
pairs of tau leptons are produced via photon--photon interaction $\gamma\gamma\rightarrow \tau^+\tau^-$ (figure~\ref{diag}a), in which a photon from the electron interacts with a photon emitted by the proton~\cite{Vermaseren:1982cz}. 
Tau leptons and tau anti--neutrinos are produced in $W$ boson decays, as illustrated in figure~\ref{diag}b~\cite{EPVEC}. 
The signature of these events is a high transverse momentum ($P_T$) isolated tau lepton, large missing transverse momentum $P^{\rm miss}_T$ due to the undetected neutrinos, and a hadronic system, typically of low $P_T$. 

\par
This paper presents a measurement of the production 
of tau lepton pairs ($\tau^+\tau^-$) and a search for events with an isolated tau lepton accompanied by large missing transverse momentum  ($\tau+P_T^\mathrm{miss}$). The measurement of $\tau^+\tau^-$  production is performed at low transverse   momentum considering both leptonic and hadronic tau decays. 
This measurement complements that of electron and muon pair production previously performed by the H1 collaboration~\cite{Aktas:2003jg,Aktas:2003sz}.
In the search for $\tau+P_T^\mathrm{miss}$ events tau decays are  identified in the hadronic channel only.
This search complements the previous  H1 measurements of the production of events with an isolated  electron~or~muon and large missing transverse momentum, which  have revealed an excess over the SM expectation of events
containing in addition a high $P_T$ hadronic system~\cite{Ahmed:1994aw,Adloff:1998aw,Andreev:2003pm}.
The ZEUS collaboration has also studied the production of events with an isolated lepton and large missing transverse momentum~\cite{Breitweg:1998aw,zeus_singletop,tauzeus}.

\par
The analysis is based on data from electron\footnote{In this paper, the name of the particle is used for both particles and anti--particles, unless otherwise stated (e.g. ``electron'' is used generically to refer to both electrons and positrons).}--proton collisions at a centre--of--mass energy of 301 or 319~GeV, recorded by the H1 experiment at HERA in the period $1994$--$2000$.
The total integrated luminosity amounts to $106$~pb$^{-1}$ for the measurement of $\tau^+\tau^-$ production
and  $115$~pb$^{-1}$ for the search for  $\tau+P_T^\mathrm{miss}$ events. 

\par
This paper is organised as follows.
The physics processes leading to tau lepton production at HERA are described in section~$2$ together with the relevant background processes.
In section~$3$ the H1 detector and the experimental conditions are briefly described. 
Particle identification is presented in section~\ref{sec:partid}.
The selection of events with tau pairs and the resulting measurements are described in section~$5$. 
The search for events with an isolated high $P_T$ tau lepton and large missing transverse momentum is presented in section~$6$. A summary is given in section~$7$.
%======================================
\section{Signal and Background Processes} 
%======================================
\label{sec:smproc}

\par
The Standard Model processes 
leading to tau lepton production in electron--proton collisions are briefly outlined in this section, together with the dominant background processes. For each process, $X$ is used to label the hadronic final state, excluding the tau decay products.

\par

The following processes, denoted henceforth as signal, lead to events containing genuine tau leptons in the final state:

\begin{itemize}

%\noindent

\item {\bf Tau pair production: $ep \rightarrow e\tau^+\tau^- X$}  \\ 
  The production of tau pairs proceeds mainly via photon--photon collisions~\cite{Vermaseren:1982cz}, as shown in figure~\ref{diag}a. 
  The proton can remain intact (elastic production, which dominates) or be  dissociated in the
  interaction (inelastic production). The incident electron is usually scattered at small angles and is often not observed in the main detector.
  Only a small  fraction of the total cross section is visible in the detector, 
  as the cross section steeply falls  with the transverse momentum $P^{\tau}_T$ of the tau leptons.
  The cross section is about 20~pb for $P^{\tau}_T$~$>$~$2$~GeV. 
  This process is modelled using the GRAPE~\cite{GRAPE} Monte Carlo (MC) generator.
  
\item {\bf Production of {\boldmath $W$} bosons: $e p \rightarrow e W X \rightarrow e \tau \nu_{\tau} X$} \\
  The dominant production mechanism for $W$ bosons in $ep$ collisions~\cite{EPVEC} is depicted in figure~\ref{diag}b. 
  The cross section is largest in the photoproduction regime (photon--proton  collisions), in which the $W$ boson usually has small transverse momentum and the  scattered electron is not observed in the main detector.
  The  $W$ production cross section at HERA is about 1~pb, 
  resulting in a cross section of about 0.1~pb for the process  $e p \rightarrow e W X  \rightarrow e \tau \nu_{\tau} X$. 
  This process is  modelled using the  EPVEC generator~\cite{EPVEC}. 
  The next--to--leading order QCD corrections to $W$ production~\cite{SPIRA} are taken into account by weighting the
  events as a function of the rapidity and transverse momentum of the $W$ boson~\cite{PRIVSPI}.

\end{itemize}

The following processes, denoted henceforth as background, do not contain genuine tau leptons but contribute to the selected samples through misidentification or mismeasurement:

\begin{itemize}

\item {\bf Neutral Current deep inelastic scattering (NC~DIS): $ep \rightarrow eX$}  \\
  The scattered electron, or  a quark or gluon that hadronises into a collimated jet of low particle multiplicity,
  may fake the hadronic tau decay signature. 
  Missing transverse  momentum may occur in NC~DIS events because of  fluctuations in the detector response
  and limited geometrical acceptance. 
  The RAPGAP~\cite{RAPGAP} generator is used to calculate this contribution to the background, including diffractive processes.

\item {\bf Photoproduction of jets: $\gamma p \rightarrow X$}  \\
  Photoproduction processes may contribute to the background if a jet in the hadronic final state 
  is misidentified as a hadronic tau decay.
  As in NC DIS events, missing transverse momentum may be measured due to fluctuations in the detector 
   response and limited geometrical acceptance. 
  The PYTHIA generator~\cite{PYTHIA}  is used to  calculate the contribution from non-diffractive hard scattering 
  photoproduction processes. 
  The contribution from diffractive photoproduction processes is simulated using the  RAPGAP generator.

\item {\bf Electron or muon pair production: $ep\rightarrow e e^+ e^- X  \; \mathrm{or} \; e\mu^+\mu^-X $ }  \\
Events with two leptons ($ee$, $\mu \mu$ or $e \mu$) may mimic
events containing $\tau^{+} \tau^{-}$ pairs in which both tau leptons decay leptonically.
The production of electron or muon pairs  may also contribute as  background 
  if one electron or muon is not identified and fakes the signature of a hadronic tau decay.
  This contribution is calculated using the GRAPE generator.

\item {\bf Charged current deep inelastic scattering (CC DIS):  $ep \rightarrow \nu X$}  \\
  CC DIS events contain genuine large missing transverse momentum due to the presence of a neutrino in the final state. A jet originating from the struck quark may fake the hadronic tau decay signature. 
  This process constitutes the main background contribution to the $\tau+P_T^\mathrm{miss}$ analysis. 
  The DJANGO~\cite{DJANGO} generator is used to calculate the background contribution from CC DIS processes.
  
\end{itemize}

\par
All generated events are passed through the full GEANT~\cite{Brun:1987ma} based simulation of the H1 apparatus and are reconstructed using the same procedure that is applied to the data. 

%======================================
\section{Experimental Conditions} 
%======================================
\label{sec:exp}

At HERA, electrons
%\footnote{In this paper, the term ``electron'' is used generically to refer to both electrons and positrons.}
of energy $E_{e} = 27.6$~GeV~collide with protons of energy $E_{p} = 820$ or $920$~GeV,
corresponding to a centre--of--mass energy $\sqrt{s}$ of $301$ or $319$~GeV, respectively.
The analysed datasets consist of $36.5$~pb$^{-1}$ of $e^{+}p$ collisions at $\sqrt{s} = 301$~GeV (taken in the period 1994-1997),
$65.1$~pb$^{-1}$ of $e^{+}p$ collisions at $\sqrt{s} = 319$~GeV (1999-2000) 
and $13.6$~pb$^{-1}$ of $e^{-}p$ collisions at $\sqrt{s} = 319$~GeV (1998-1999), corresponding to a total integrated luminosity of $115.2$~pb$^{-1}$.
 The data recorded in the period 1994-1995 are not included for the measurement of tau pair production,
 reducing the integrated luminosity to $106.0$~pb$^{-1}$ for this analysis.

\par

A detailed description of the H1 detector can be found in \cite{H1detector}.  
Only the components  essential to the present analysis are described here. 
A right handed cartesian  coordinate system is used with the origin at the nominal primary $ep$ interaction vertex. 
The proton beam direction defines the $z$ axis. The polar angles $\theta$ and transverse momenta $P_T$ of all particles 
are defined with respect to this axis.  The azimuthal angle $\phi$ defines the particle direction in the transverse plane. The pseudorapidity  is defined as $\eta=-\ln {\tan {\frac{\theta}{2}}}$.

\par

The inner tracking system contains the central ($20^\circ < \theta <
160^\circ$) and forward ($7^\circ$~$<$~$\theta$~$<$~$25^\circ$) drift
chambers, and the proportional chambers which are employed for triggering purposes. It is used to determine the position of the interaction
vertex and to measure the trajectories of charged particles.  
Particle transverse momenta and charges are determined from the curvature of the trajectories in a solenoidal magnetic field of $1.15$~T.

\par

Hadronic final state particles as well as electrons and photons are
absorbed in the highly segmented liquid argon~(LAr) 
calorimeter~\cite{h1cal} ($ 4^\circ < \theta < 154^\circ$) which is  $20$ to $30$
radiation lengths deep depending on the polar angle.
The hadronic section of the LAr calorimeter is $5$
to $8$ hadronic interaction lengths deep. Electromagnetic shower energies are measured
with a precision of $\sigma (E) / E = 12\% / \sqrt{E/\mathrm{GeV}}
\oplus 1\%$ and hadronic shower  energies with $\sigma (E) / E = 50\% /
\sqrt{E/\mathrm{GeV}} \oplus 2\%$, as determined in test beam
measurements~\cite{h1testbeam}.  
In the backward region ($153^\circ < \theta < 178^\circ$), 
the LAr calorimeter is complemented\footnote{In 1994, a lead-scintillator calorimeter~\cite{h1bemc} was used instead.} by a  lead--scintillating fibre spaghetti 
calorimeter~(SpaCal)~\cite{h1spacal}. 
The LAr and SpaCal  calorimeters are enclosed within the superconducting coil and 
are surrounded by an iron  return yoke which is instrumented with streamer tubes 
to allow for the detection of  muons
in the range $4^\circ < \theta < 171^\circ$. 
\par
Dissociated proton states may be 
measured at small polar angles by a set of detectors in the forward direction: the PLUG, a sandwich calorimeter constructed from copper  plates and silicon
counters ($0.6^\circ < \theta < 3.5^\circ$),
 the proton remnant tagger
(PRT), an array of  scintillators ($0.06^\circ$~$<$~$\theta$~$<$~$0.17^\circ$), and the drift chambers of the forward
muon  detector (FMD)~\cite{h1fmd} ($3^\circ$~$<$~$\theta$~$<$~$17^\circ$). 

\par
The
luminosity is determined from the rate of Bethe-Heitler events $ep\rightarrow e\gamma p$, where the photon is detected in a calorimeter located downstream of the interaction point. 

\par
The events selected in this analysis are triggered by detecting  
electromagnetic clusters in the LAr or SpaCal calorimeter (electron trigger), 
by measuring a large missing transverse  momentum in the LAr calorimeter ($P^{\rm miss}_T$ trigger), or by using hits in the muon detectors   combined with central tracker signals (muon trigger).
In the kinematic range of this  analysis, events containing a pair of tau leptons are triggered with an efficiency of   about $55\%$ ($35\%$)
if one  tau lepton decays leptonically (both tau leptons decay hadronically).
Events containing single tau leptons and missing transverse momentum above $12$~GeV ($25$~GeV)
are triggered with an efficiency of about $50\%$ ($85\%$).

%======================================
\section{Particle Identification} 
%======================================
\label{sec:partid}

\subsection{Identification of electrons}
\label{sec:elid}

An electron candidate is defined by the presence of a compact and isolated electromagnetic cluster
of energy deposits in the LAr or SpaCal calorimeter.
The kinematics of the electron candidate are measured from the calorimeter cluster.
Among the charged tracks reconstructed in the event,  the track with the lowest extrapolated distance of closest approach to the cluster ($\delta$) is associated to the electron if it satisfies the condition $\delta<12$~cm. In this case, the azimuthal angle $\phi$ and the charge of the electron are measured from the associated track.
Additional energy within a cone of radius $0.5$ in the pseudorapidity--azimuth ($\eta$--$\phi$) plane
around the electron candidate is required to be less than $3\%$ of the energy attributed to the electron candidate.
The electron identification efficiency is estimated using NC DIS events and is greater than $95\%$ for an electron energy above $5$~GeV.

\subsection{Identification of muons}
\label{sec:muid}

A muon candidate is identified by a track in the inner tracking system
associated with a track segment or an energy deposit in the instrumented iron.
The muon momentum is measured from the track curvature in the solenoidal magnetic field.
The rejection of hadrons traversing the calorimeter and reaching the muon detectors is improved
by requiring that the muon candidate deposits less  than $5$~GeV around its extrapolated track in the LAr calorimeter
within a cylinder of radius $35$~cm in the electromagnetic and $75$~cm in the hadronic section.
The efficiency to identify muons is estimated using elastic $\gamma \gamma \rightarrow \mu^{+} \mu^{-}$ events
and is greater than $85\%$ in the energy range considered in this analysis.

\subsection{Reconstruction of the hadronic final state}
\label{sec:hfs}
The hadronic final state (HFS) is measured~\cite{hadroo2} by combining energy measurements from the calorimeter 
with charged particle momenta measured by the inner tracker. 
Identified isolated electrons or muons are excluded from the HFS.
The hadronic energy scale is calibrated by comparing the
transverse momentum of well measured electrons to that of the HFS in NC DIS events~\cite{benjamin}. 
Jets with  $P_T > 2$~GeV are reconstructed from HFS particles using an inclusive $k_T$ algorithm~\cite{Ellis:tq} in the $P_T$ recombination scheme with a separation parameter set to one.

\subsection{Identification of tau decays}
\label{sec:tauid}

Leptonic tau decays (branching ratio $35\%$~\cite{Eidelman:2004wy}) are identified by detecting an electron or a muon as described above. Hadronic tau decays typically produce low multiplicity, collimated, hadronic jets, henceforth denoted as $\tau$-jets. 
Depending on the number of charged hadrons produced, the 
hadronic decay modes are summed up in two categories, referred to as ``1-prong'' (one charged hadron, branching ratio $49\%$) and ``3-prong'' (three charged hadrons,  branching ratio $14\%$).  The branching ratio for decays into more than three charged hadrons is 
small (about $2\%$) and such decays are not considered in the present analysis.

\par
Two different algorithms to identify hadronic tau decays, applied to jets reconstructed in the angular range $20^\circ<\theta <120^\circ$, have been developed and are described below. The measurement of tau lepton pair production, in which the tau leptons generally have low momentum, requires an optimal background rejection. 
In contrast, the search for tau leptons produced in $W$ decays at high $P_T$ uses an algorithm that maximises the identification efficiency, since the signal cross section is low and the background is less severe.

\subsubsection*{ A:  Neural Network based identification algorithm optimised for low \boldmath{$P_T$} taus} 
For the measurement of the tau lepton pair production,  
an algorithm has been developed that uses multiple neural
networks to discriminate between $\tau$--jets and the background from  electrons, muons or hadronic jets.  A detailed description of the algorithm can be found in~\cite{thesisCristianVeelken}. 
The algorithm is implemented in two steps.

\par
In the first step, 1--prong (3--prong) candidates for hadronic
tau decays are preselected by requiring exactly one (three) well reconstructed track(s) in the drift chambers within a cone (``$\tau$-cone'') around the jet axis. The opening angle of the $\tau$--cone varies between 5$^\circ$ and 30$^\circ$ depending on the jet momentum, with 
smaller angles at higher momentum due to the larger Lorentz boost in the direction of the tau candidate. The tracks are required to be not associated to identified electrons and muons and the scalar sum of their transverse momenta is required to be larger than $2$~GeV.
The fine granularity of the LAr calorimeter is used  to match extrapolated tracks with energy deposits in the calorimeter and to reconstruct additional neutral particles associated to the tau candidate from unmatched energy deposits in the $\tau$--cone. The sum of the four--vectors of the tracks and of the neutral particles defines the $\tau$--jet four--vector.  If all associated  tracks have a well  measured charge,
% ($|\kappa/\Delta \kappa | > 2$, where $\kappa$ denotes the curvature of the track and $\Delta \kappa$ the corresponding uncertainty), 
the  charge of the tau candidate is reconstructed as the sum of the charges of the associated tracks. 

\par
In the second step of the algorithm, various variables related to the particle multiplicity and collimation of the $\tau$--jet candidate are used.
 The set of variables includes: the multiplicity 
 of the neutral clusters within the $\tau$--cone; 
the invariant masses calculated from clusters, from tracks and from charged and neutral particles in the $\tau$--cone; the difference in energy measured from tracks and from clusters; the distance in $\eta-\phi$ between tracks and clusters; the first and second transverse moments of the distribution of energy deposits in the calorimeter with respect to the jet axis; the sum of energy deposits detected in an extended cone of radius $1.0$ in $\eta$-$\phi$ around the $\tau$--cone.
Two neural networks (NN) using
    these variables are separately trained using MC simulations to
    identify 1-prong and 3-prong tau decay modes, the output of which is denoted by $L^\mathrm{1-prong}$ and $L^\mathrm{3-prong}$, respectively.
Their numerical value varies between zero and one and is used to discriminate between tau candidates (close to one) and hadronic jets (close to zero). Depending on the track multiplicity of the jet,
    the output of one of these NNs is used to select tau candidates.
By requiring $L^\mathrm{1-prong}>0.75$ (for 1--prong candidates) or $L^\mathrm{3-prong}>0.75$ (for 3--prong candidates),  the efficiency of this algorithm to identify hadronic tau lepton decays in $\tau^+\tau^-$ events is about $50\%$, as calculated from MC simulations of tau decays with visible energy in the range considered in this analysis.
The probability for hadronic jets to be misidentified as hadronic 1--prong (3--prong) tau decays is $0.5\%$ ($4\%$). 
\par
The signature of hadronic 1--prong tau decays may also be faked by unidentified electrons and muons. Two additional neural networks are trained to veto such cases. 
The output of these NNs, denoted by $L_\mathrm{veto}^{e}$ and $L_\mathrm{veto}^{\mu}$ respectively, is expected to be close to one for tau decays and close to zero for electrons or muons that fail the identification criteria described in sections~4.1~and~4.2.

\subsubsection*{B: Identification algorithm optimised for high \boldmath{$P_T$} taus}
A different approach to tau identification is used to search for events with a high momentum tau lepton and large $P^{\rm miss}_T$. A high identification efficiency and  a sufficient level of  background rejection are achieved by 
requiring a collimated jet, containing only one charged particle and  isolated with respect to other tracks and jets within a cone of radius $1.0$ in 
$\eta$--$\phi$. 
The identification of hadronic $\tau$-decays is therefore restricted  to hadronic 1--prong decay modes.
\par
The collimation of a jet is measured by the jet radius $R_{jet}$, 
defined as the energy weighted average distance in $\eta$--$\phi$  between the jet axis and all HFS particles $h$ (neutral and charged) in the jet:
\begin{equation}
R_{jet} = \frac{1}{E_{jet}}\sum_h E_h \sqrt{
  \Delta\eta(jet,h)^2+\Delta\phi(jet,h)^2 } \;.
 \label{eq:rjet}
 \end{equation}
Jets with one track and  $R_{jet} < 0.12$ are selected as tau candidates. The four--vector of the tau candidate is taken to be that of the jet. For jets with $P_T>7$~GeV, this identification procedure results in an efficiency of about $80\%$ to identify 1--prong hadronic decays of tau leptons resulting from decays of $W$ bosons. The misidentification probability for hadronic jets is less than $1\%$.

%======================================
\section{Production of  \boldmath{$\tau^{+} \tau^{-}$ }Pairs}
%======================================

\subsection{Event selection}
\label{sec:taupSelec}

The decay modes investigated in the present analysis are 
 classified as {\em leptonic} when both taus decay leptonically (branching ratio $6.2\%$, excluding decays to same flavour leptons), \linebreak
{\em semi--leptonic}   when one tau decays hadronically and the second leptonically ($45\%$) and {\em hadronic}  when both taus decay hadronically ($42\%$).
The case where both tau leptons decay to charged leptons of the same flavour ($ee$ or $\mu\mu$ final state) is not included in the present analysis, 
as the separation of the  $\tau^{+} \tau^{-}$ signal from electron or muon pair production is difficult. 

\par
Leptonic and hadronic tau decays are identified as described in the previous section and by applying in addition the following selection criteria. The isolation of electrons and muons is measured by the distance in the $\eta-\phi$ plane to the closest hadronic jet ($D_{\rm jet}$)
and to the closest track ($D_{\rm track}$).

\begin{itemize}
\item $e$ candidates are reconstructed in the polar angular region $20^{\circ} < \theta_{e} < 140^{\circ}$ 
      and are required to have an energy above 5 GeV and a transverse momentum above 3~GeV. 
      They must be isolated from jets with  $D_{\rm jet} >1.0$.

\item $\mu$ candidates are reconstructed in the polar angular region $20^{\circ} < \theta_{\mu} < 140^{\circ}$
      and are required to have a transverse momentum above 2~GeV. 
      They must  be isolated from other tracks and jets with $D_{\rm track}>0.5$ and $D_{\rm jet} >1.0$, respectively.

\item $\tau$--jet candidates of transverse momenta above 2~GeV 
      are reconstructed in the polar angular region $20^{\circ} < \theta_{\tau} < 120^{\circ}$,
      using algorithm {\bf A} as described in section~\ref{sec:tauid}.
      The output of the neural network is required to satisfy  $L^\mathrm{1-prong}>0.75$ or $L^\mathrm{3-prong}>0.75$.
\end{itemize}

\par
Events with two tau candidates are selected. If the charge of both tau candidates is measured, events with two candidates of the same charge are rejected. 

\par 
In order to avoid significant background contributions from NC DIS and 
$\gamma p$ processes,
the  analysis is restricted to elastic $ep\rightarrow ep\tau^+ \tau^-$ production. 
Inelastic  events are vetoed by requiring no extra track or energy deposit above  the noise level
in the main detector in addition to those associated to the decay products of both tau leptons and a possible additional electron. 
Furthermore, no significant activity should be observed in the forward detectors  (PLUG, FMD and PRT). The requirements applied~\cite{laycockthesis} ensure that the proton remains intact or dissociates into a low mass state.
Remaining background originates mainly from elastic lepton pair production  and from diffractive NC DIS or $\gamma p$ processes.

\par
Electron and muon pair production processes ($\gamma \gamma \rightarrow e^+e^-$ and $\gamma \gamma \rightarrow \mu^+\mu^-$)
constitute a background to tau pair production in semi--leptonic decay modes when one of the leptons is correctly identified as $e$ or $\mu$ 
while the second lepton fails to be identified by the algorithms described in sections~\ref{sec:elid}~and~\ref{sec:muid}
and fakes the signature of a 1--prong hadronic tau decay.
In order to reject this background, $L_\mathrm{veto}^{e}> 0.75$ ($L_\mathrm{veto}^{\mu}>0.75$) is required
for 1--prong tau jet candidates selected in the $e + \tau$--jet ($\mu + \tau$--jet) channel.

\par
To further reduce the background from NC DIS processes,  where the scattered electron is selected as a candidate for a $\tau\rightarrow e$ decay
and the struck quark fragments into a collimated jet of low particle multiplicity that fakes the $\tau$--jet signature, the longitudinal momentum balance calculated as
\begin{equation}
E-P_z = \sum_{i} E_i(1-\cos \theta_i)\; 
\label{eq:epz}
\end{equation}
is employed, where $E_i$ and $\theta_i$ denote the energy and polar angle of each particle detected in the event.
For events in which only momentum in the proton direction is undetected,
$E-P_z$ is equal to twice the energy $E_e$ of the electron beam, i.e. $55.2$~GeV.  For events containing tau leptons, $E-P_z$ values well below $55.2$~GeV are expected since the neutrinos produced in the tau decays are not detected 
and, most of the time, the scattered beam electron escapes down the beam pipe. 
By requiring $E-P_z < 50$~GeV when the detected electron has the same charge as the beam lepton,
the NC DIS background is rejected to a large extent.  The $E-P_z$ cut is also applied to the leptonic channel and rejects muon pair--production events, for which the scattered electron is detected together with one produced muon while the second muon escapes in the forward direction. If a second electron is detected in the event, this condition is not applied.
\renewcommand{\arraystretch}{1.5}
\begin{table}[h]
\begin{center}
\tablesize
\begin{tabular}{|l|c|c|c|c|}
\hline
\multicolumn{5}{|c|}{\bf Selection of \boldmath{$\tau^+ \tau^-$} Events} \\
\hline
% Channel & $e \ + \ \mu$ & $e \ + \ \tau$-jet & $\mu \ + \ \tau$-jet & $\tau$--jet + $\tau$--jet \\
 {\bf Decay Channel} & {\bf Leptonic} & \multicolumn{2}{|c|}{\bf Semi--leptonic} & {\bf Hadronic} \\
\hline 
\hline
\multirow{3}{18mm}{$\tau$~Signatures} & $e$                         & $e$                       & 
                                        $\mu$                     & $\tau$--jet                     \\

                                      & $\mu$                          & $\tau$--jet                       & 
                                        $\tau$--jet                    & $\tau$--jet                     \\ \cline{2-5}
                         
                                       & \multicolumn{4}{|c|}{Events with two same charge tau candidates rejected } \\
\hline
\multirow{2}{18mm}{Elastic Production} & \multicolumn{4}{|c|}{No additional tracks, no additional clusters in the LAr or SpaCal calorimeter,} \\
                                       & \multicolumn{4}{|c|}{no activity in forward detectors above noise level} \\

\hline
\multirow{2}{18mm}{Background Reduction} &                        & $L_\mathrm{veto}^{e}>0.75^{(1)}$                      & 
                                         $L_\mathrm{veto}^{\mu}>0.75^{(1)}$                    & % $P_{T}^\mathrm{miss} > 2$~GeV          
					 \\ \cline{2-5}

                                      & \multicolumn{2}{|c|}{$E-P_z<50$~GeV$^{(2)}$}                 
                                                         &          &  %$P_{T}^{miss} > 2$~GeV       
							 
							 \\ 
\hline
\hline
\multicolumn{5}{|l|}{ $^{(1)}$ applied only to 1--prong $\tau$--jet candidates} \\
\multicolumn{5}{|l|} {$^{(2)}$ applied only if the electron associated with the tau decay has the same charge as the beam lepton} \\ 
\multicolumn{5}{|l|} {$^{ }$ \;\;\; and no second electron is detected} \\
\hline

\end{tabular}
\end{center}
\vspace{-5mm}
\begin{center}
\caption{ Selection criteria for elastic $\tau^{+} \tau^{-}$ events in the leptonic ($e +  \mu$), semi--leptonic ($e + \tau$--jet, $\mu  + \tau$--jet)
                      and hadronic ($\tau$--jet $+$ $\tau$--jet) decay modes of the $\tau$ lepton pair.}
\label{tableTauPairEventSelection}
\end{center}
\end{table}
\par
The selection criteria for $\tau^{+} \tau^{-}$ events are summarised in table~\ref{tableTauPairEventSelection}. 
With these selection criteria,  $1.2\%$ of the elastic $\gamma \gamma \rightarrow \tau^{+} \tau^{-}$ events in which both tau leptons satisfy $P_T^\tau>2$~GeV and $20^\circ<\theta_\tau<140^\circ$
are selected.
The efficiency is limited by the fact that the energy of the detected tau decay products is significantly lower than the tau lepton energy, since the neutrinos escape detection.

\subsection{Background studies}
\label{sec:taupBgStudy}
Due to the aforementioned elasticity requirements, the remaining background from  
NC DIS and $\gamma p$ processes consists mainly of exclusive diffractive  events, 
for which the validity of the
resolved pomeron model~\cite{Ingelman:1984ns} as implemented in the RAPGAP program is questionable. Hence the RAPGAP prediction for NC DIS ($\gamma p$)  is normalised to the number of events observed in a control sample in which an electron and a jet (two jets) are selected in the $P_T$ and $\theta$ ranges of the analysis, and where the elastic requirements are applied. It has been verified that the shapes of the observed kinematic distributions are reasonably well described by RAPGAP.
\par
Furthermore, to check that the probability for an electron, a  muon or a hadronic jet to be misidentified as a $\tau$--jet candidate is well described by the MC simulation, several event samples are studied in which  the contribution of individual background processes is enhanced.
These event samples are selected in a phase space  
similar to that of the $\tau^{+} \tau^{-}$ event sample, 
requiring  the conditions for elastic production to be fullfilled. In order to enhance the background component in the control samples, 
 the condition $L^\mathrm{1-prong}>0.75$ or $L^\mathrm{3-prong}>0.75$ 
 on the NN outputs for $\tau$--jet candidates is not applied.

The following samples are employed:
\begin{itemize}
\item {\bf  \boldmath{$ \gamma \gamma \rightarrow e^{+} e^{-}$} control sample} \\
A $\gamma \gamma \rightarrow e^{+} e^{-}$ dominated event sample 
is defined  by selecting events with one electron and one 1--prong $\tau$--jet candidate.  No veto condition  on $L_\mathrm{veto}^{e}$  is  applied. Events with two same charge tau candidates are rejected.
\item {\bf \boldmath{$ \gamma \gamma \rightarrow \mu^{+} \mu^{-}$} control sample} \\
A $\gamma \gamma \rightarrow \mu^{+} \mu^{-}$ enriched event sample is defined by selecting events with one muon and one 1--prong $\tau$--jet candidate.
No veto condition  on $L_\mathrm{veto}^{\mu}$  is  applied. Events with two same charge tau candidates are rejected.
\item {\bf \boldmath{$\gamma p$} control sample} \\
A $\gamma p$ enriched event sample is defined
by selecting  events containing two jets with one, two or three tracks.
%$\tau$--jet candidates. 
If both jets are 1--prong tau candidates, the requirements  $L^e_{veto} > 0.75$ and $L^\mu_{veto} > 0.75$ are applied to suppress the contributions from $\gamma \gamma \rightarrow e^{+} e^{-}$ and $\gamma \gamma \rightarrow \mu^{+} \mu^{-}$ processes. In order to reduce the contribution from NC DIS process, only events with $E-P_z<45$~GeV are accepted.

\item {\bf NC DIS control sample} \\
A NC DIS enriched event sample is defined by selecting
events with one electron and one $\tau$--jet candidate.
If the $\tau$--jet is a 1--prong candidate, $L^e_\mathrm{veto} > 0.75$ is required 
in order to suppress the contribution from $\gamma \gamma \rightarrow e^{+} e^{-}$ processes.
\end{itemize}

The selection criteria of the background control samples are summarised in table~\ref{tab:ttcontrol}, in which the numbers of events obtained from the data and the MC simulation are also given.
\renewcommand{\arraystretch}{1.5}
\begin{table}[h]
\begin{center}
\tablesize
\begin{tabular}{|l|c|c|c|c|}
\hline
\multicolumn{5}{|c|}{\bf \boldmath{$\tau^+ \tau^-$} Background Control Samples} \\
\hline
\multirow{2}{18mm}{\bf Control samples} & \boldmath{$e^+e^-$} & \boldmath{${\mu^+\mu^-}$} & \boldmath{$\gamma p$} &{\bf NC DIS}\\
 & & & & \\
\hline
\hline
\multirow{3}{18mm}{Signatures} & $e$ & $\mu$ &  two jets & $e$    \\
                                      &  1--prong $\tau$--jet &1--prong $\tau$--jet  & with $\le 3$ tracks  & $\tau$--jet    \\  \cline{2-5}
                         
                                       & \multicolumn{2}{|c|}{Events with two same charge} & & \\
                                       & \multicolumn{2}{|c|}{ tau candidates rejected } & & \\
\hline
\multirow{2}{18mm}{Elastic Production} & \multicolumn{4}{|c|}{No additional tracks, no additional clusters in the LAr or SpaCal calorimeter,} \\
                                       & \multicolumn{4}{|c|}{no activity in forward detectors above noise level} \\
\hline
\multirow{2}{18mm}{Background Reduction } & & & $L_\mathrm{veto}^{e}>0.75^{(1)}$ & $L_\mathrm{veto}^{e}>0.75^{(2)}$ \\
                                          & & & $L_\mathrm{veto}^{\mu}>0.75^{(1)}$ &  \\ 
                                          & & & $E-P_z<45$~GeV&  \\ 
\hline 
\hline                                        
H1 Data    & $115$           & $20$          & $29$            & $29$\\
Total SM   & $133.1\pm 19.5$ & $14.1\pm 1.4$ & $24.9 \pm 10.0$ & $32.4 \pm 6.3$\\ 
 & ($95$\% $e^+e^-$)  &  ($50$\% $\mu^+\mu^-$)  &  ($79$\% $\gamma p$)&  ($62$\% NC DIS)  \\   

\hline
\multicolumn{5}{|l|}{$^{(1)}$ applied only if both jets are 1--prong tau candidates } \\
\multicolumn{5}{|l|}{$^{(2)}$ applied only to 1--prong $\tau$--jet candidates} \\
\hline
\end{tabular}
\end{center}
\vspace{-5mm}
\begin{center}
\caption{Selection criteria for the background control samples in which each background contribution is individually enhanced. The numbers of observed and expected events are also shown. The dominant contribution to the total SM expectation is indicated as a percentage in the last row.
         Here, the tau candidates are not required to satisfy the condition $L^\mathrm{1-prong}>0.75$ or $L^\mathrm{3-prong}>0.75$. 
	 }
\label{tab:ttcontrol}
\end{center}
\end{table}
\par
\begin{figure}[h]
\setlength{\unitlength}{1mm}
\begin{center}
\begin{picture}(150,189)(0,0)

\put(0.,0.)   {\epsfig{file=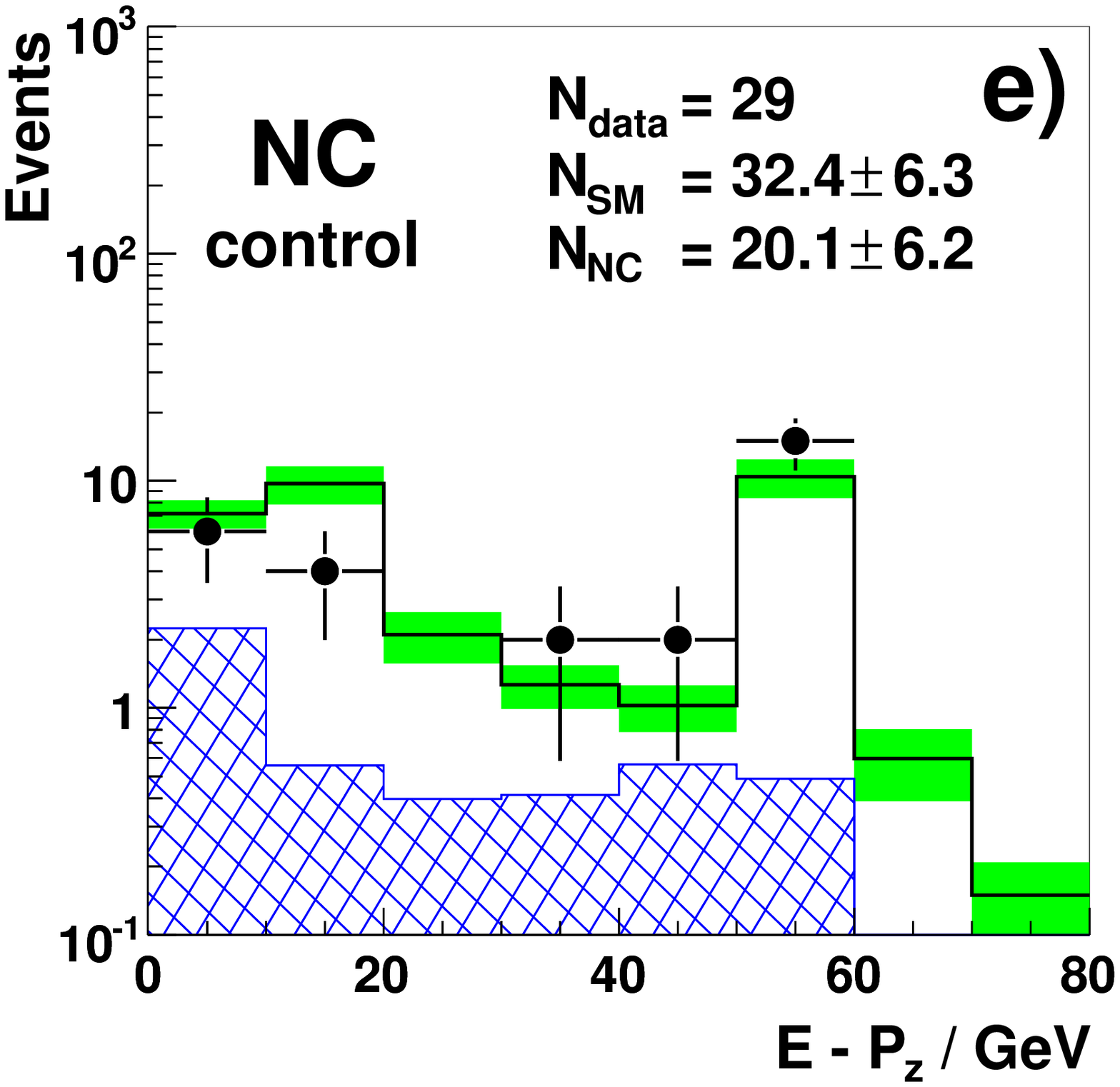,bbllx=15,bblly=0,bburx=560,bbury=560,width=7.0cm,clip=}}
\put(70.,10.) {\epsfig{file=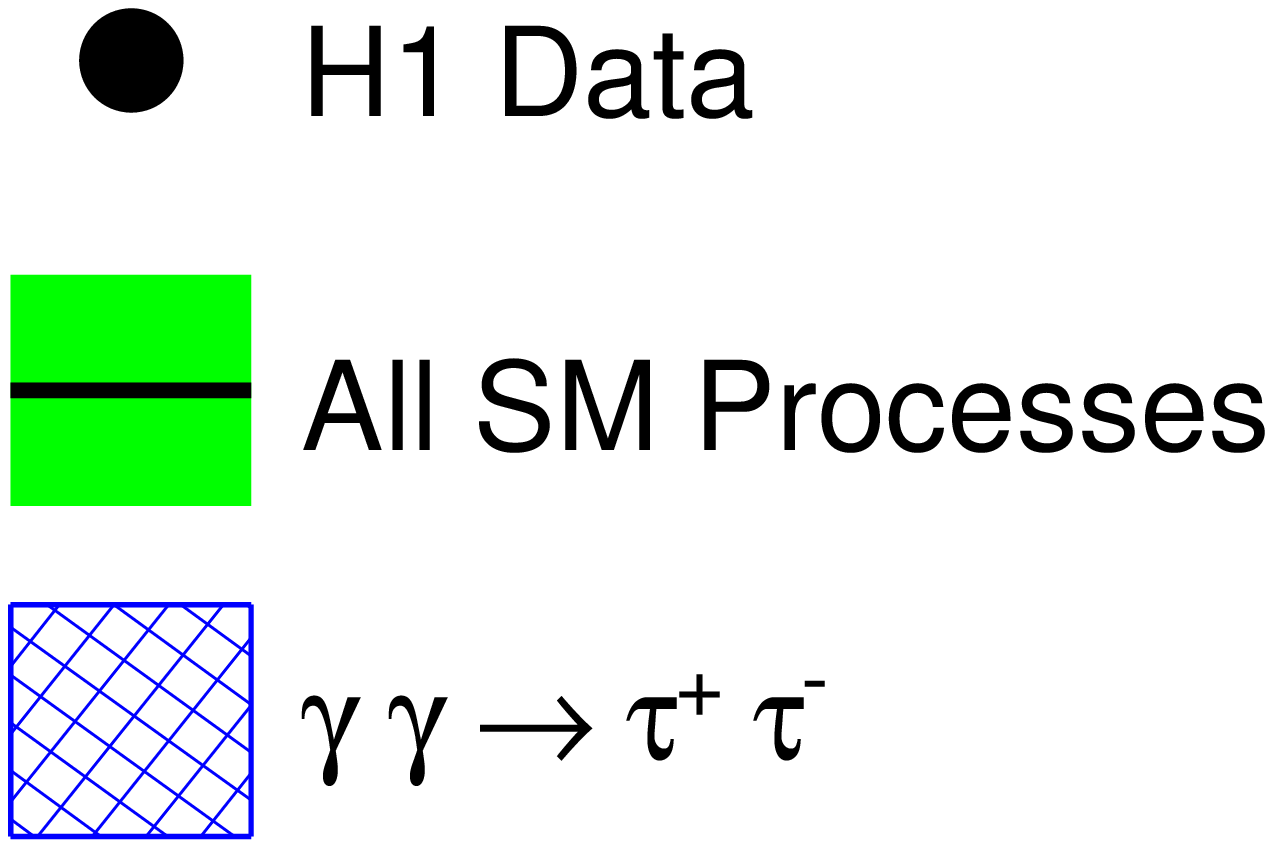,bbllx=15,bblly=0,bburx=560,bbury=560,width=7.0cm,clip=}}
\put(0.,63.)  {\epsfig{file=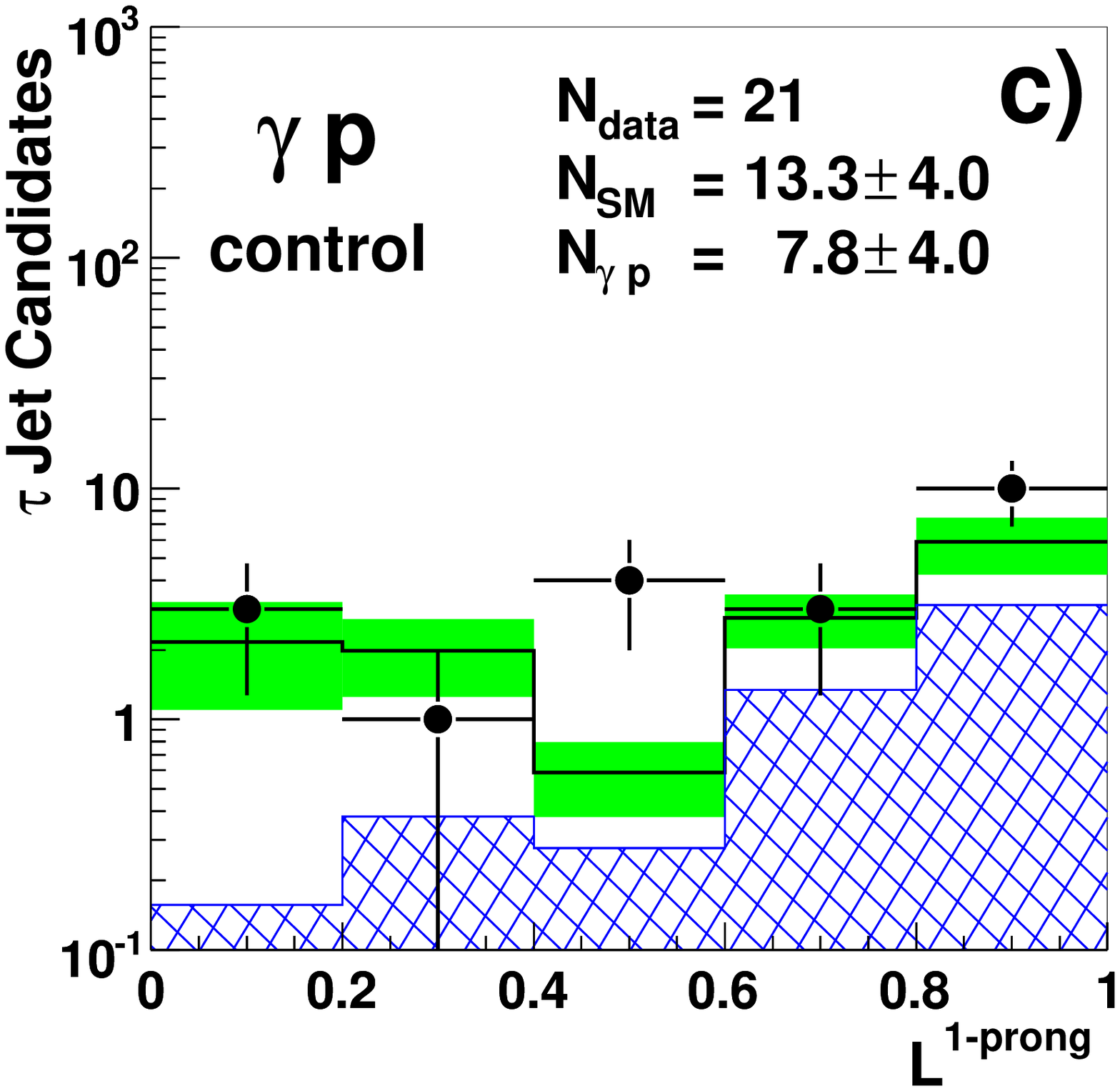,bbllx=15,bblly=0,bburx=560,bbury=560,width=7.0cm,clip=}}
\put(70.,63.) {\epsfig{file=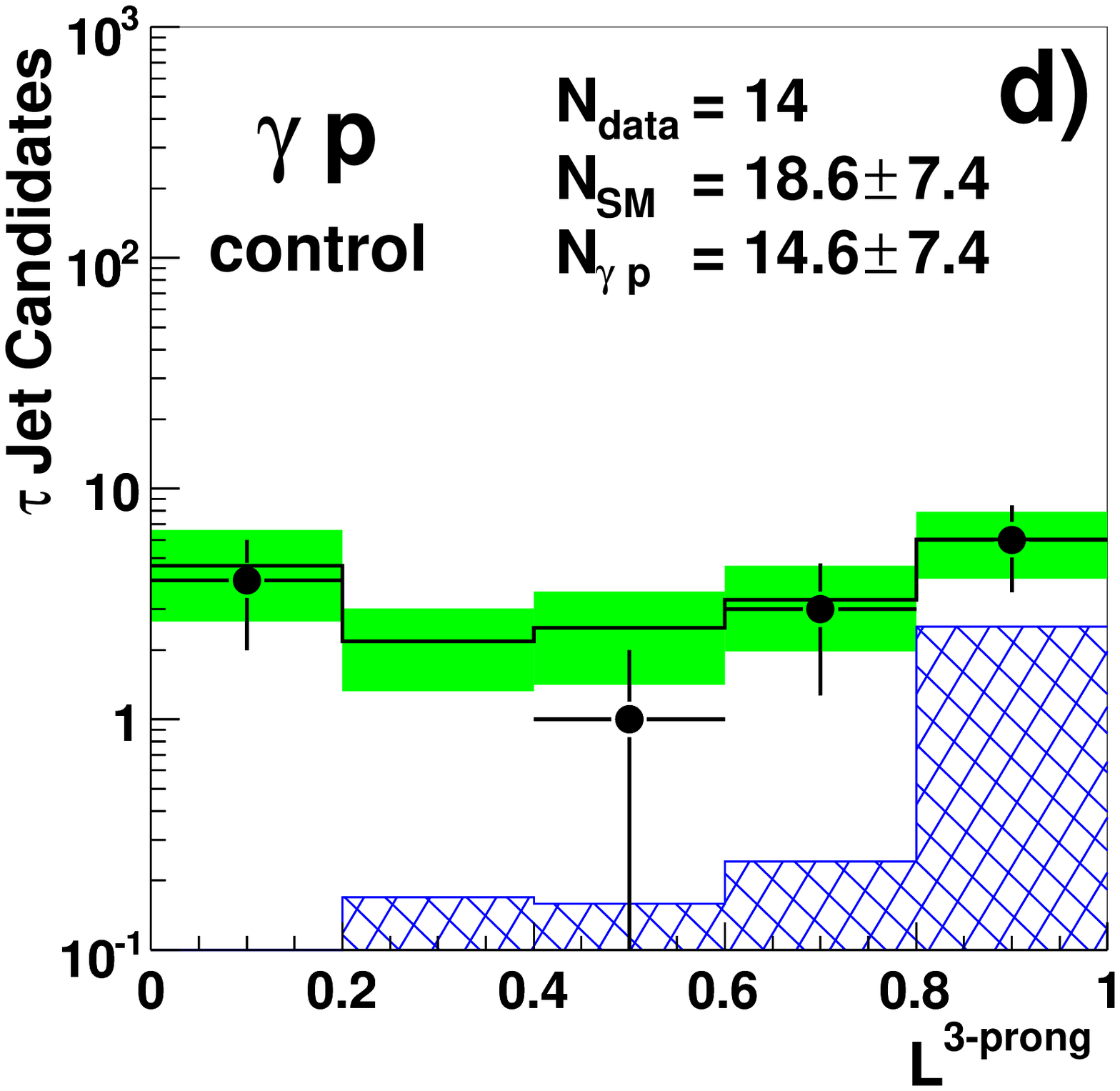,bbllx=15,bblly=0,bburx=560,bbury=560,width=7.0cm,clip=}}
\put(0.,126.) {\epsfig{file=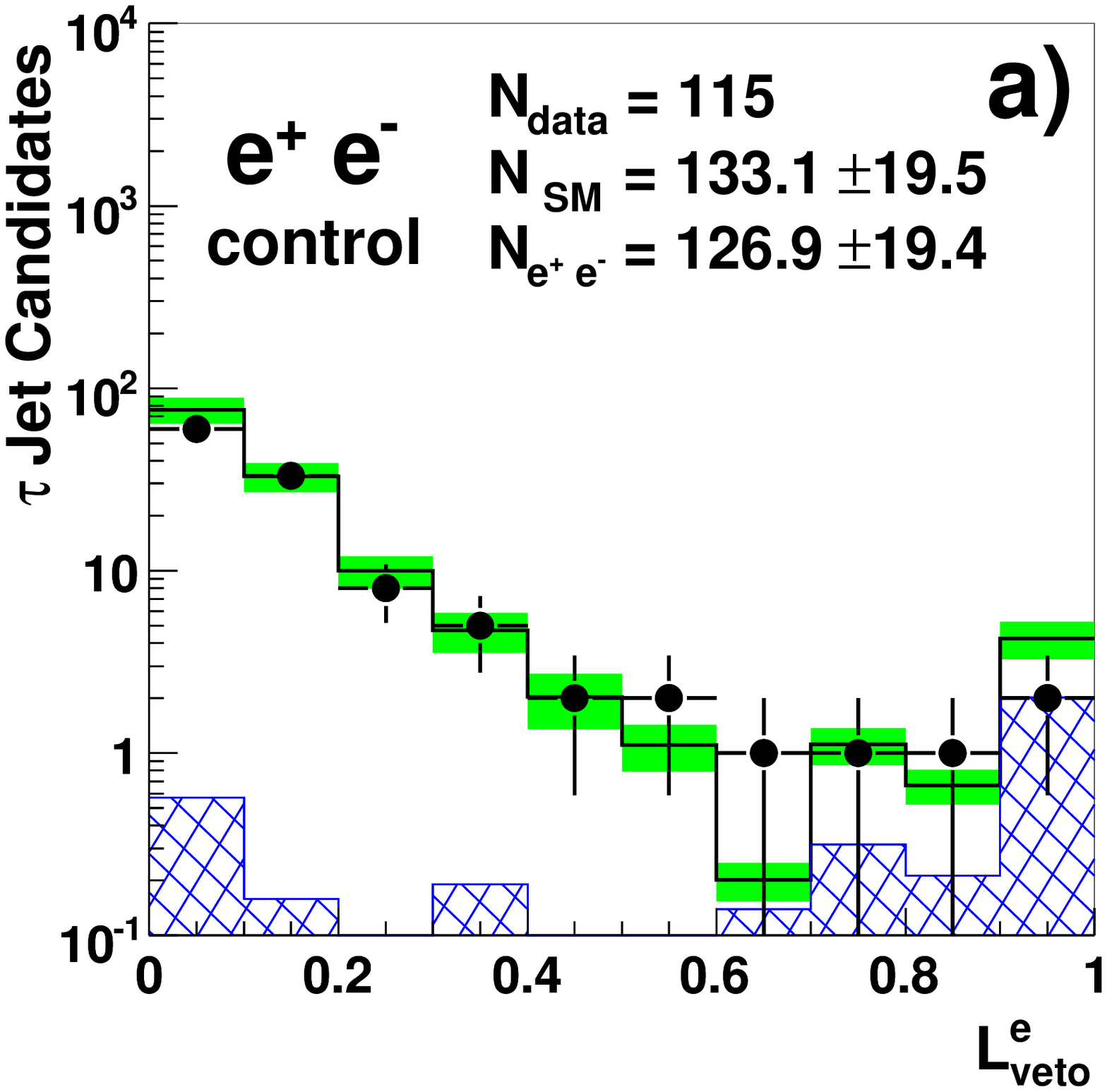,bbllx=15,bblly=0,bburx=560,bbury=560,width=7.0cm,clip=}}
\put(70.,126.){\epsfig{file=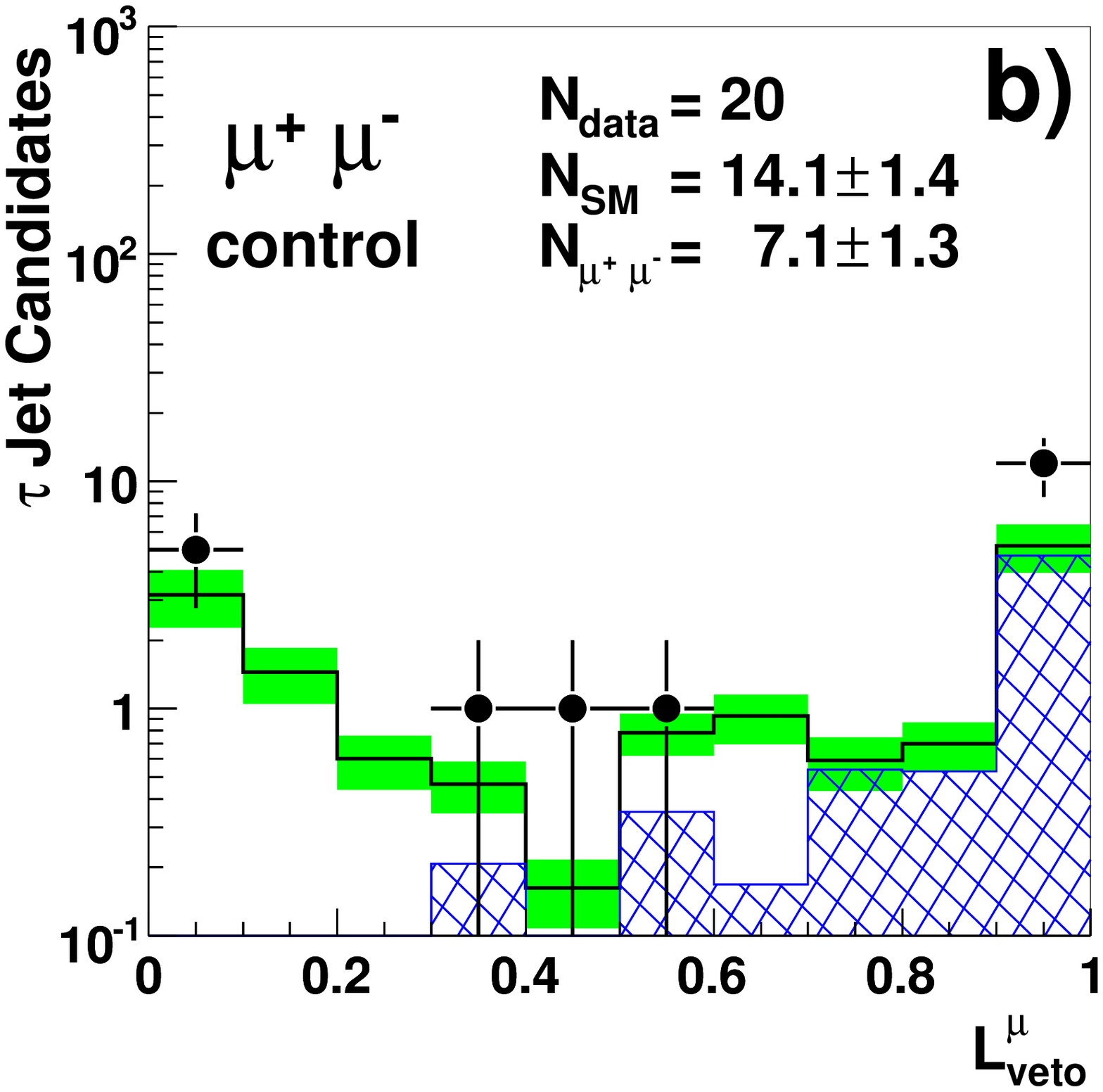,bbllx=15,bblly=0,bburx=560,bbury=560,width=7.0cm,clip=}}
\end{picture}
\caption{Distributions of {\bf a)} the likelihood $L^e_\mathrm{veto} $ for $\tau$--jet candidates 
         in the $\gamma \gamma \rightarrow e^{+} e^{-}$ control sample, {\bf b)}
         the likelihood $L^\mu_\mathrm{veto} $ for $\tau$--jet candidates 
         in the $\gamma \gamma \rightarrow \mu^{+} \mu^{-}$ control sample,
         {\bf c)} and {\bf d)} the NN outputs $L^\mathrm{1-prong}$  and $L^\mathrm{3-prong}$  for 1--prong and 3--prong $\tau$--jet candidates 
	  in the $\gamma p$ control sample, and {\bf e)}
         the longitudinal momentum balance $E - P_{z}$ in the NC DIS control sample. In each figure the open histogram shows the total SM expectation and the shaded band its uncertainty. The contribution from tau pair production is shown as the hatched histogram.}
\label{figureTauPairBackgroundEnrichedSamples}
\end{center}
\end{figure}

The distributions of all quantities used in the selection of the $\tau^{+} \tau^{-}$ event sample
are well described by the MC simulation both in shape and normalisation in the control samples.
Examples of these distributions are shown in figure~\ref{figureTauPairBackgroundEnrichedSamples}. The distributions of the electron and muon rejection discriminators $L_\mathrm{veto}^e$ and $L_\mathrm{veto}^\mu$ are shown in figures~\ref{figureTauPairBackgroundEnrichedSamples}a~and~\ref{figureTauPairBackgroundEnrichedSamples}b for the control samples where most of the tau candidates are unidentified electrons and muons, respectively. The distribution of the NN output  $L^\mathrm{1-prong}$ ($L^\mathrm{3-prong}$) is shown in figure~\ref{figureTauPairBackgroundEnrichedSamples}c~(\ref{figureTauPairBackgroundEnrichedSamples}d) for 1--prong (3--prong) jets in the  $\gamma p$ control sample which is enriched in hadronic jets. In figures~\ref{figureTauPairBackgroundEnrichedSamples}a--d the contribution from tau pair production populates the region close to one, while the background accumulates at values close to zero, as expected. 
Finally, the distribution of the $E-P_z$ variable is shown in figure~\ref{figureTauPairBackgroundEnrichedSamples}e for the NC DIS control sample.
The agreement between data and simulation in the control samples 
shows that the background contribution as well as the experimental efficiencies are modelled by the MC simulation within the attributed systematic uncertainties described in the next section.

\subsection{Systematic uncertainties}
\label{sec:tpSystematics}
In this section, the systematic uncertainties associated with the 
measurement of elastic $\tau^{+} \tau^{-}$ production 
are discussed.
The uncertainties on the signal expectation 
are determined by varying the following experimental quantities  by $\pm 1$ standard deviation.
\begin{itemize}
%%%%%%%%%%%%%%%%%%%%%%%%%%%%%%%%%%%%%%%%%%%%%%%%%%%
\item {\bf Electron identification and reconstruction } \\
In the kinematic region considered in the $\tau^{+} \tau^{-}$ analysis,
the uncertainty on the electron identification efficiency is $5\%$ ($2\%$ when the electron energy is above $10$~GeV).
The electron energy scale uncertainty is estimated to be $3\%$.
The uncertainties on the measurement of the electron angles $\theta$ and $\phi$ are $3$~mrad and $1$~mrad, respectively.
%%%%%%%%%%%%%%%%%%%%%%%%%%%%%%%%%%%%%%%%%%%%%%%%%%%
\item {\bf Muon identification and reconstruction  } \\
The uncertainty on the muon identification efficiency is $5\%$.
The relative scale uncertainty on the muon momentum  is conservatively taken to be $5\%$.
The uncertainties on the measurements of the muon angles $\theta$ and $\phi$ are $3$~mrad and $1$~mrad, respectively.
%%%%%%%%%%%%%%%%%%%%%%%%%%%%%%%%%%%%%%%%%%%%%%%%%%%
\item {\bf Identification and reconstruction of hadronic tau decays  } \\
For each of the charged tracks associated to the $\tau$--jet, a reconstruction efficiency uncertainty of $3\%$ is assigned.
The efficiency to identify hadronic $\tau$ decays with the NN algorithm {\bf A} has an additional uncertainty of $10\%$~\cite{thesisCristianVeelken}, 
estimated by comparing different simulations of shower development in the LAr calorimeter. 
It has been verified, using a large statistics sample of hadronic jets from inclusive CC DIS and NC DIS samples, that the output of both NNs is well described by the MC simulation. 
The uncertainties arising from the modelling of tau decays are negligible, as estimated  by comparing the results obtained using either the PYTHIA~\cite{Sjostrand:2003wg} or  TAUOLA~\cite{Tauola} programs to simulate the decays of the tau leptons generated with GRAPE.
The energy of the neutral clusters of $\tau$--jet candidates has a relative uncertainty of $4\%$, corresponding to the hadronic energy scale uncertainty.
The uncertainties on the measurements of the $\tau$--jet angles $\theta$ and $\phi$ are estimated to be $10$~mrad and $5$~mrad, respectively.
%%%%%%%%%%%%%%%%%%%%%%%%%%%%%%%%%%%%%%%%%%%%%%%%%%%
\item {\bf Elastic event selection } \\
The efficiency with which elastic events are selected and proton dissociative events are rejected
depends on the noise level in the LAr and on the performance of the forward detectors~\cite{laycockthesis}. Its uncertainty does not exceed $3\%$.
%%%%%%%%%%%%%%%%%%%%%%%%%%%%%%%%%%%%%%%%%%%%%%%%%%%
\item {\bf Triggering } \\
The trigger efficiency is studied using elastic $e^{+} e^{-}$ and $\mu^{+} \mu^{-}$ events
and diffractive $\gamma p$ di-jet events in a phase space similar to that of the  $\tau^{+} \tau^{-}$  analysis.
The uncertainty on the trigger efficiency depends on the region in $\theta$ in which the tau candidates are detected: central ($\theta>30^\circ$) or forward ($\theta<30^\circ$). The
 uncertainty is $10\%$ if both tau candidates are in the central region, $20\%$ if one tau candidate is detected in the forward and the other in the central region and $30\%$ if both tau candidates are detected in the forward region.
%%%%%%%%%%%%%%%%%%%%%%%%%%%%%%%%%%%%%%%%%%%%%%%%%%%
\item {\bf Luminosity }  \\
The luminosity of the analysed datasets is measured with an uncertainty of $1.5\%$.
%%%%%%%%%%%%%%%%%%%%%%%%%%%%%%%%%%%%%%%%%%%%%%%%%%%
\end{itemize}

The individual effects of the above experimental uncertainties are combined in quadrature,
yielding a total uncertainty of $21\%$ on the signal expectation. The largest contributions to this uncertainty arise from systematics attributed to the  tau identification procedure ($15\%$) and to the trigger efficiency ($12\%$).

\par
Contributions from background processes, modelled using the generators described in section $2$,
are attributed relative systematic uncertainties of $50\%$ ($\gamma p$), $30\%$ (NC DIS), \linebreak $15\%$ ($\gamma \gamma \rightarrow e^+ e^-$, $\gamma \gamma \rightarrow \mu^+ \mu^-$), 
estimated from the level of agreement observed between the MC simulation and the data in the background enhanced control samples 
described in section \ref{sec:taupBgStudy}.

\subsection{Results}

\begin{figure}[h]
\setlength{\unitlength}{1mm}
\begin{center}
% \begin{picture}(150,150)(0,0)
% \put(0,150){\mbox{\epsfig{file = fig3.eps, angle=270, width=17cm }}}
% \end{picture}
\epsfig{file = 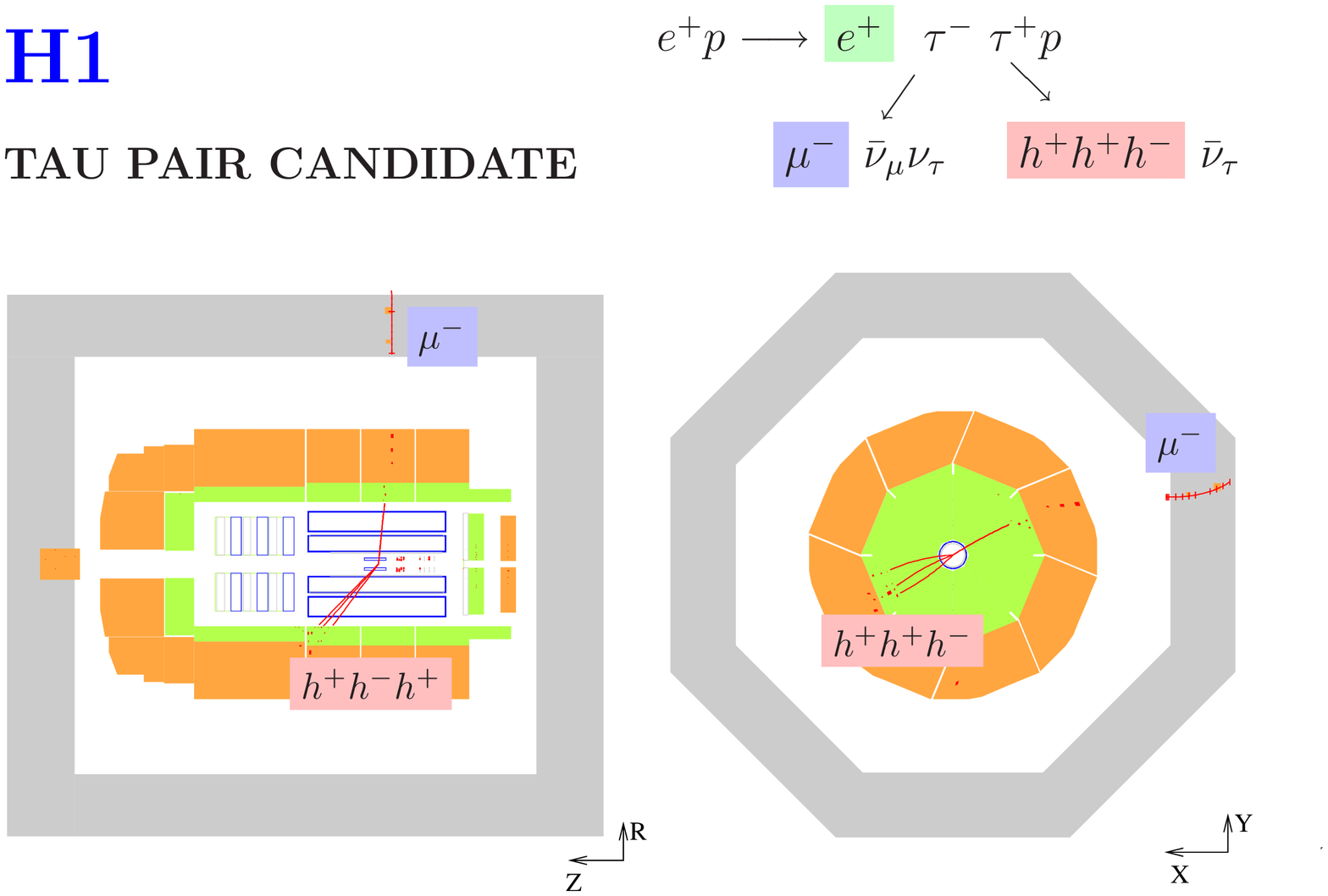, angle=270, width=17cm }
\end{center}
\caption{ A tau pair candidate event with one tau lepton decaying
  leptonically to a muon and the other tau lepton decaying to three
  charged hadrons (3-prong topology). }
\label{event}
\end{figure}

\begin{figure}[h]
\setlength{\unitlength}{1mm}
\begin{center}
\begin{picture}(150,135)(0,0)
\put(0.,0.)   {\epsfig{file=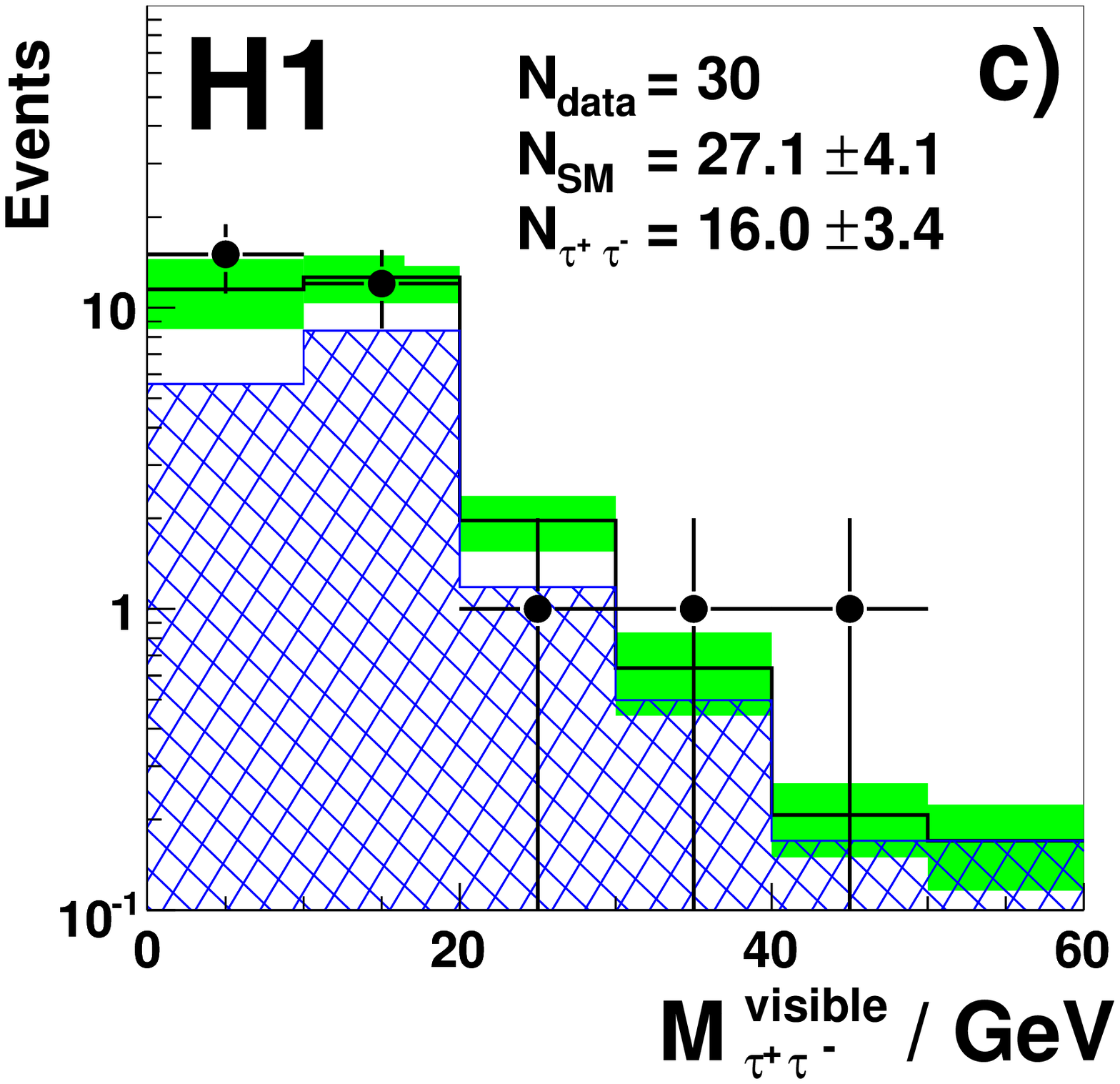,bbllx=15,bblly=0,bburx=560,bbury=560,width=7.7cm,clip=}}
\put(85.,15.) {\epsfig{file=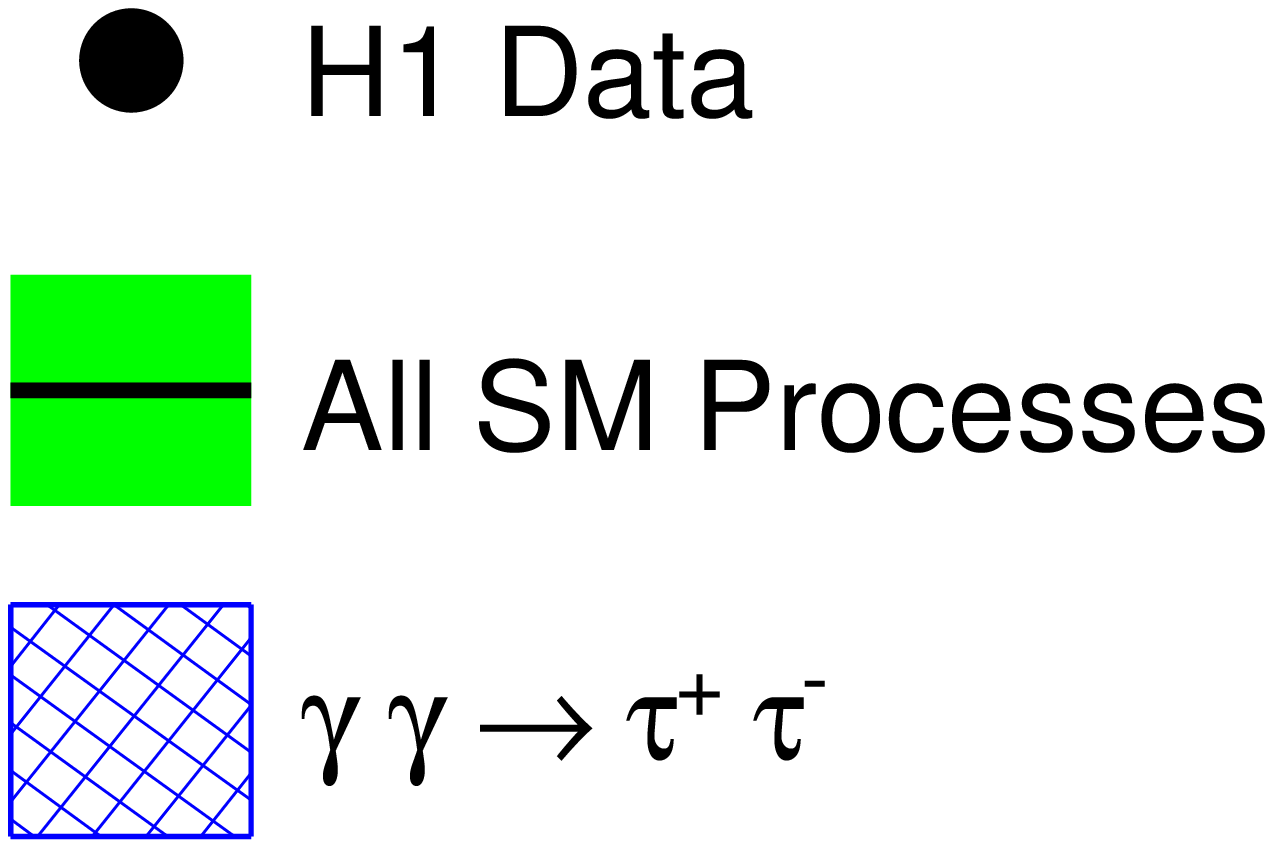,bbllx=15,bblly=0,bburx=560,bbury=560,width=6.8cm,clip=}}
\put(0.,70.)  {\epsfig{file=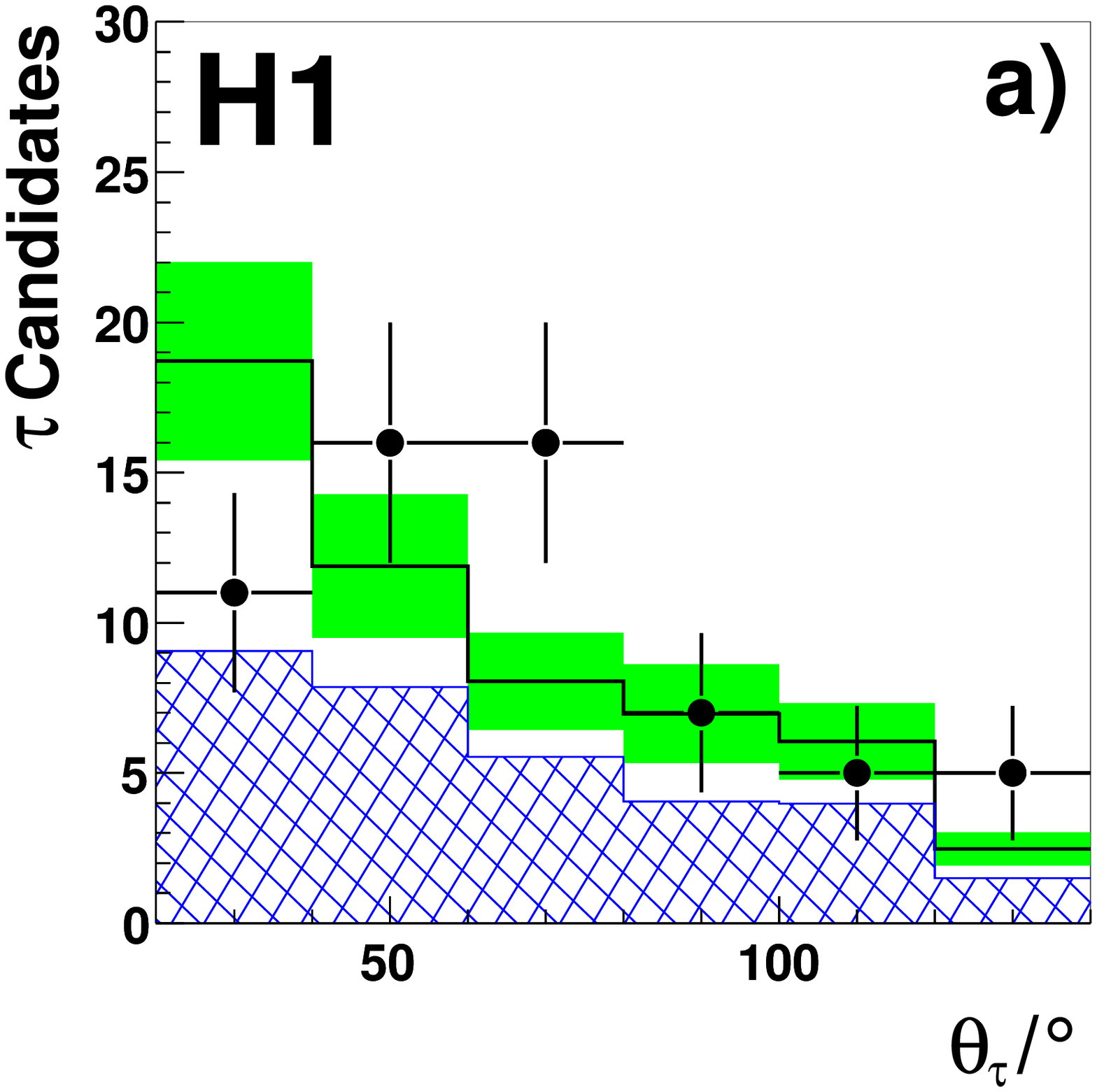,bbllx=15,bblly=0,bburx=560,bbury=560,width=7.7cm,clip=}}
\put(75.,70.) {\epsfig{file=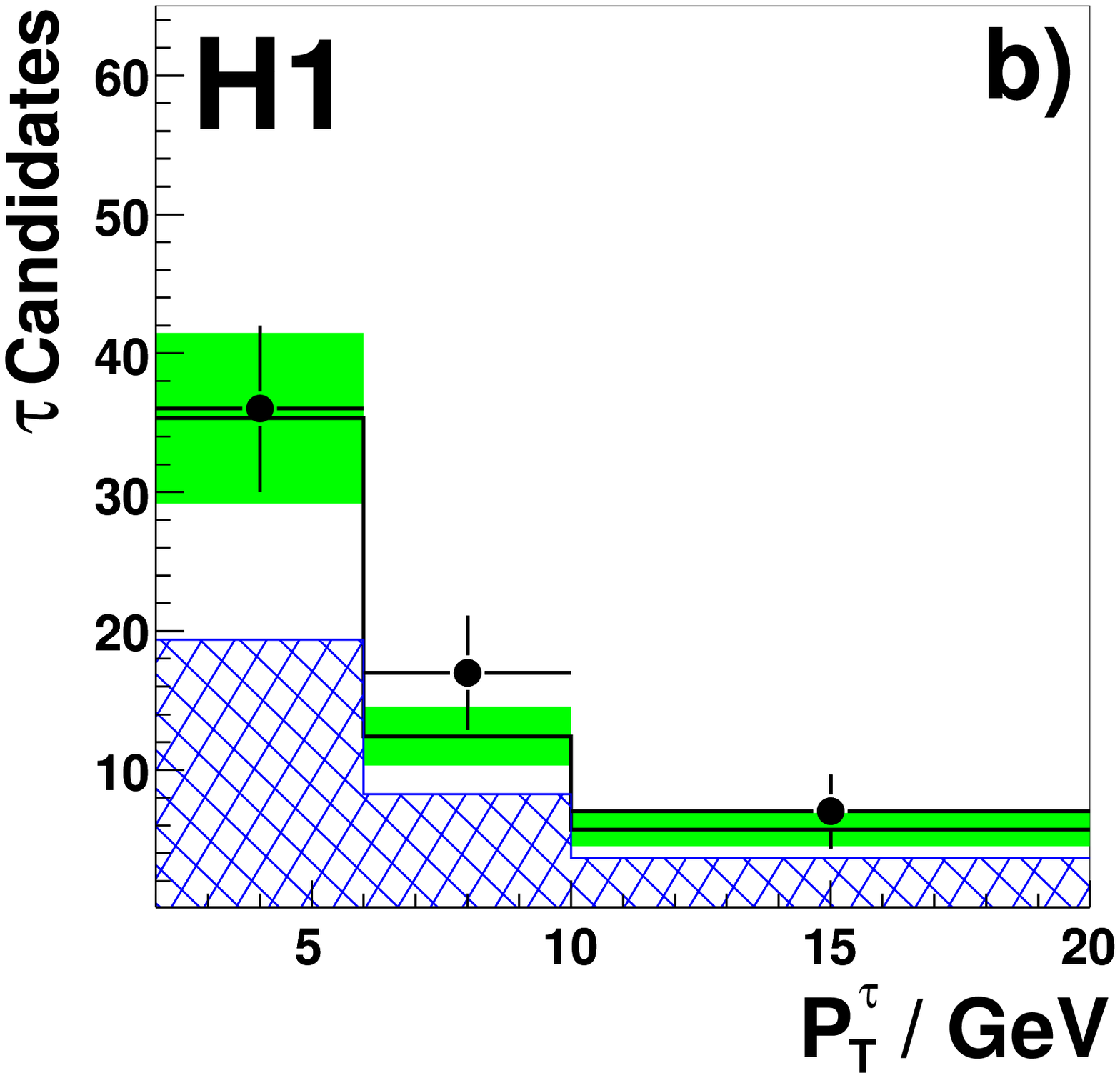,bbllx=15,bblly=0,bburx=560,bbury=560,width=7.7cm,clip=}}

\end{picture}

\caption{Distributions of {\bf a)} the polar angle 
         and {\bf b)} the transverse momentum of the reconstructed $\tau$ lepton candidates  in the $\tau^{+} \tau^{-}$ event sample. The invariant mass of the tau candidate pair is shown in {\bf c)}.  The selected events enter  distributions {\bf a)} and {\bf b)} twice.
	 In each figure the open histogram shows the total SM expectation and the shaded band its uncertainty. The contribution from tau pair production is shown as the hatched histogram.}
\label{figureTauPairResults1}
\end{center}
\end{figure}
In total, $30$ $\tau^{+} \tau^{-}$ candidate events are selected,
in agreement with the SM expectation of $27.1 \pm 4.1$ events, of which $16.0 \pm 3.4$ are expected from $\gamma \gamma \rightarrow \tau^{+} \tau^{-}$ signal processes. The signal expectation is  dominated ($85\%$) by the elastic production component.  The lepton pair production, NC DIS and $\gamma p$ processes contribute with similar rates to the background expectation. The numbers of observed and expected events in the four analysed channels are shown in table~\ref{tab:taupaires}. 
An event selected in the semi--leptonic channel  is displayed in figure~\ref{event}.
\renewcommand{\arraystretch}{1.2}
\begin{table}[h]
\begin{center}
\begin{tabular}{|c||c|c|c|c||c|} \hline
\multicolumn{6}{|c|}{\bf  \boldmath{$\tau^+ \tau^-$} Results} \\
\hline
\hline
{\bf Decay Channel} & {\bf Leptonic} & \multicolumn{2}{|c|}{\bf Semi--leptonic} & {\bf Hadronic} & {\bf Total} \\ 
\cline{3-4}
 & $e$ $\mu$ & $e$ $\tau$--jet & $\mu$ $\tau$--jet & $\tau$--jet $\tau$--jet & \\ 
\hline
H1 Data & $7$ & $2$ & $10$ & $11$ & $30$ \\
SM & $2.9\pm 0.4$ & $6.3\pm 0.9$ & $7.0\pm 1.3$ & $11.0 \pm 2.0$ & $27.1\pm 4.1$\\
$\tau^+\tau^-$ & $56\%$& $47\%$ & $85\%$ & $50\%$ & $59\%$ \\  
\hline
\end{tabular}
\caption{Number of selected events and SM prediction for the $\tau^+\tau^-$ analysis. 
         The expected relative contribution of the $\tau^+\tau^-$ process to the SM prediction is also shown.} 
\label{tab:taupaires}
\end{center}
\end{table}

\par
The distributions of the polar angle and of the transverse momentum of the tau candidates, together with the visible invariant mass, are shown in figure~\ref{figureTauPairResults1}.
The measured  distributions are compatible with the SM expectations.
As expected for $\gamma \gamma \rightarrow \tau^{+} \tau^{-}$ processes, most tau decay products are detected
with relatively small transverse momenta. \label{sec:tpXsection}

\par
Using the selected sample,  a measurement of the cross section for the elastic production of $\tau^{+} \tau^{-}$ pairs is performed in the kinematic region defined by $20^{\circ} < \theta_{\tau} < 140^{\circ}$ and  $P_{T}^{\tau} > 2$~GeV.
For this measurement, the data samples collected at $\sqrt{s}=301$~GeV and $319$~GeV are combined,
taking into account their respective luminosities.
Assuming a linear dependence of the cross section on the proton beam energy, as predicted by the SM,
the measured cross section corresponds to an effective centre--of--mass energy of $\sqrt{s} = 314$~GeV.
The cross section is calculated using the formula:
\begin{equation}
\sigma = \frac{N_\mathrm{data} - N_\mathrm{bgr}}{\mathcal{L} \cdot A},
\label{equationCrossSectionCalculation}
\end{equation}
where $N_\mathrm{data}$ is the number of observed events, 
$N_\mathrm{bgr}$ the expected contribution from background processes, $\mathcal{L}$  the total integrated luminosity and $A$  the signal acceptance. 
The contribution from inelastic $\gamma \gamma \rightarrow \tau^{+} \tau^{-}$ processes is included in the background expectation.
The signal acceptance $A$ is calculated using the GRAPE generator, as the ratio of the number of events accepted at reconstructed level to the number of events generated in the defined phase space. It accounts for the selection and trigger efficiencies and for the differences in momentum between the original $\tau$ leptons and the detected decay products. 

\par

The measured cross section for elastic tau pair production $ep \rightarrow ep \tau^{+} \tau^{-}$ integrated over the phase space defined above is 
$13.6 \pm 4.4 \pm 3.7~\text{pb}$
where the first error is statistical and the second systematic.
The result is in agreement with the SM expectation of $11.2 \pm 0.3~\text{pb}$,
calculated using the GRAPE generator.

%======================================
\section{Production of High \boldmath{$P_T$} Tau Leptons in Events with Large Missing Transverse Momentum} 
%======================================
\renewcommand{\arraystretch}{1.2}  

\subsection{Event selection}
\label{sec:eventsel}

Events containing an isolated  tau lepton and large missing transverse momentum are selected with a procedure similar to that used in the analysis of events with an isolated electron or muon and large missing transverse momentum~\cite{Andreev:2003pm}. 
The tau leptons are identified using hadronic decays only, as  the leptonic tau decays lead to final states which cannot be distinguished from those studied in~\cite{Andreev:2003pm}.

\par
The event selection is performed in two steps.
In the first step, the net transverse momentum 
  reconstructed from all particles (electrons, muons and hadrons)  $P^{\rm miss}_{T}$, is required to be above $12$~GeV.
In order to ensure uniform trigger conditions, the 
 net transverse momentum measured from
  all energy deposits detected in the calorimeter, $P^{\rm calo}_{T}$, is required to be above $12$~GeV.
  The reconstructed $P^{\rm miss}_{T}$ is approximately equal to $P^{\rm calo}_{T}$
  except for events containing muons in the final state.
In order to exploit further the event topology in the transverse plane, the  variable $V_{ap}/V_{p}$ is employed, defined as the ratio of the anti--parallel to parallel projections of all energy deposits in the calorimeter
  with respect to the direction of  $P^{\rm calo}_{T}$~\cite{Adloff:2000qj}.
Events with genuine missing transverse momentum
  are  in general reconstructed with  $V_{ap}/V_{p}$ values close to zero and large values of $P^{\rm calo}_{T}$,   whereas background events from NC DIS and  $\gamma p$ processes are intrinsically balanced, with larger values of $V_{ap}/V_{p}$ and low values of $P^{\rm calo}_{T}$.
  Only events with $V_{ap}/V_{p}$ below $0.5$ are accepted. For events with  $P^{\rm  calo}_{T}$ below $25$~GeV, a stricter condition of $V_{ap}/V_{p}<0.15$ is applied. The background from NC DIS is further reduced by requiring  \linebreak $E-P_z$~$<$~$50$~GeV.
Additionally, the events are required to contain at least one isolated hadronic jet with transverse momentum above 7 GeV in the central region of the detector\linebreak $20^\circ$~$<$~$\theta_{jet}$~$<$~$120^\circ$. The isolation  is characterised by the distance of the jet  in $\eta-\phi$  
to the nearest hadronic jet ($D_{\rm jet}>1.0$) and the nearest track  not belonging to the jet ($D_{\rm track}>1.0$).

\par
In the second selection step, the isolated jets are required to pass the tau identification criteria 
of algorithm {\bf B} described in section~\ref{sec:tauid}. A jet is accepted if it is narrow ($R_{jet}<0.12$) and contains exactly one charged track ($N^{jet}_{tracks}=1$).
In order to remove background from hadronic jets containing mostly neutral particles, 
this track is required to have a transverse momentum $P_T^{track}$ greater than $5$~GeV. If more than one isolated jet satisfies these requirements, the one with the highest $P_T$ is considered as the tau candidate.
In order to further reduce the background from intrinsically balanced events, 
in which the $P_T^\mathrm{miss}$ is due to a mismeasurement, 
the acoplanarity $\Delta \phi$, defined as the angle in the transverse plane between the $\tau$--jet candidate  and the hadronic system excluding the tau candidate ($X$), is required to be below $170^{\circ}$. This criterium 
removes events in which the $\tau$--jet candidate and the rest of the hadronic system are back--to--back, 
as is typical for NC DIS and $\gamma p$ events.

\par 
A summary of all selection criteria is presented in table~\ref{taucuts}. 
Using these selection criteria, SM $W \rightarrow \tau \nu$ events are selected with an overall efficiency of $8\%$.
In comparison to $W$ decays into electrons or muons~\cite{Andreev:2003pm},
$W\rightarrow \tau \nu $ decays are selected with  a significantly lower efficiency, mainly due to the branching ratio ($49\%$) for hadronic 1--prong tau decays and to the more restricted polar angular range of this analysis. 
\renewcommand{\arraystretch}{1.5}
\begin{table}[hhh]
\begin{center}
\begin{tabular}{|l|r l|}
\hline
\multicolumn{3}{|c|}{\bf Selection of \boldmath{$\tau + P^{\rm miss}_T$} Events} \\
\hline \hline
\multirow{4}{20mm}{$P_T^\mathrm{miss}$ + Isolated Jet Preselection}& $P_T^\mathrm{calo}$ & $>\;\;$~$12$ GeV \\
 & $P_T^\mathrm{miss}$ & $>\;\;$~$12$ GeV \\
 & $E-P_z$ & $<\;\;$~$50$ GeV \\
 & $V_{ap}/V_{p}$ & $<\;\;0.5$ \;\;($<0.15 \;\mathrm{for}\; P_T^\mathrm{calo}<25$~GeV) \\ 
 & $N_{jets}$ & $>\;\;$~$1$ \\
 & $P^{jet}_T$ & $>\;\;$~$7$ GeV \\
 & $20^\circ$ & $<\;\;\theta_{jet}$~$\;\;<\; 120^\circ$ \\ 
 & $D_{track}$ & $>\;\; 1.0$  \\
 & $D_{jet}$ & $>\;\; 1.0$  \\
\hline
\multirow{5}{18mm}{Final $\tau+P_T^\mathrm{miss}$ Selection }& $N^{jet}_{tracks}$ & $ =\;\; 1$  \\
& $P_T^{track}$ & $>\;\; 5$ GeV \\
& $R_{jet}$             & $<\;\; 0.12$ \\
& $\Delta \phi$ & $< \;\;170^{\circ} $ \\ 
\hline
\end{tabular}
\caption{Selection criteria for events containing an isolated $\tau$ lepton and large $P^{\rm miss}_T$.}
\label{taucuts}
\end{center}
\end{table}

\subsection{Background studies}
\label{sec:backgstudy}

After applying the selection criteria, the main background is expected to occur from events with genuine missing transverse momentum produced by CC DIS processes, 
in which a narrow jet with low track multiplicity fakes the tau signature. Additional background arises from NC DIS and $\gamma p$ processes, 
which have a much larger cross section than the CC DIS process and lead to events that contain hadronic jets but no genuine $P_T^\mathrm{miss}$. 
The modelling of the CC DIS, NC DIS and $\gamma p$ backgrounds is verified using samples in which the contribution of each background process is 
enhanced. The selection criteria defining the background control samples are listed in table~\ref{backgrcut},
together with the observed number of events and the SM expectation.

\par
The CC DIS background control sample is selected using the ``$P_T^\mathrm{miss}$ + isolated jet'' preselection described in section~\ref{sec:eventsel}. The NC DIS and $\gamma p$ samples are selected in a complementary phase space at large $E-P_z$ and large $V_{ap}/V_{p}$, respectively. In these background samples the tight selection criteria of tau candidates are not applied and only isolated  jets,
as defined in table~\ref{taucuts}, are considered.

\par
\begin{figure}[hhh]
\setlength{\unitlength}{1mm}
\begin{center}
\begin{picture}(150,165)(0,0)

\put(0,110){\mbox{\epsfig{file = 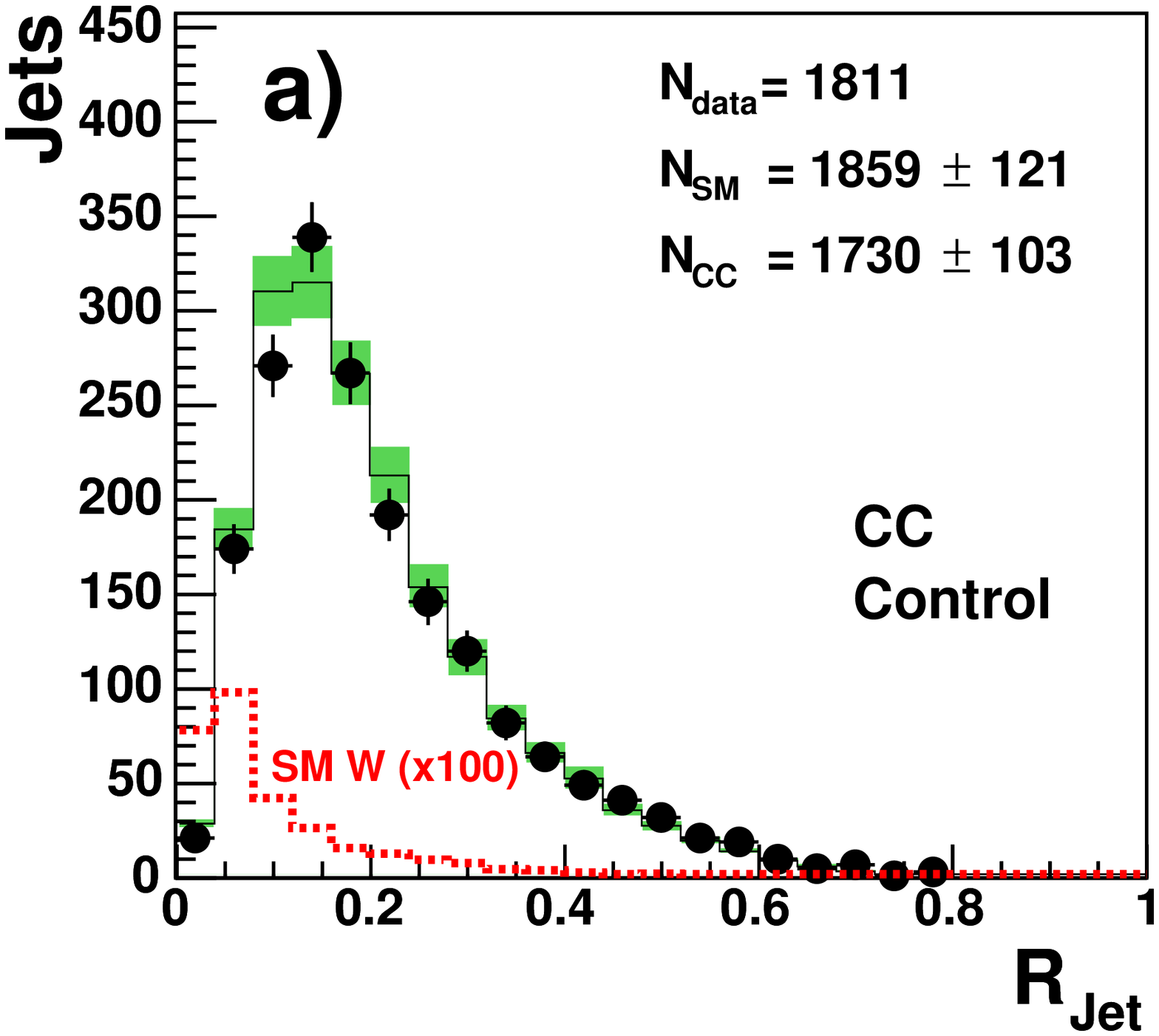,
                          bbllx=-2,bblly=1,bburx=577,bbury=574, angle=0, clip=, width=6.5cm}}}
\put(70,110){\mbox{\epsfig{file = 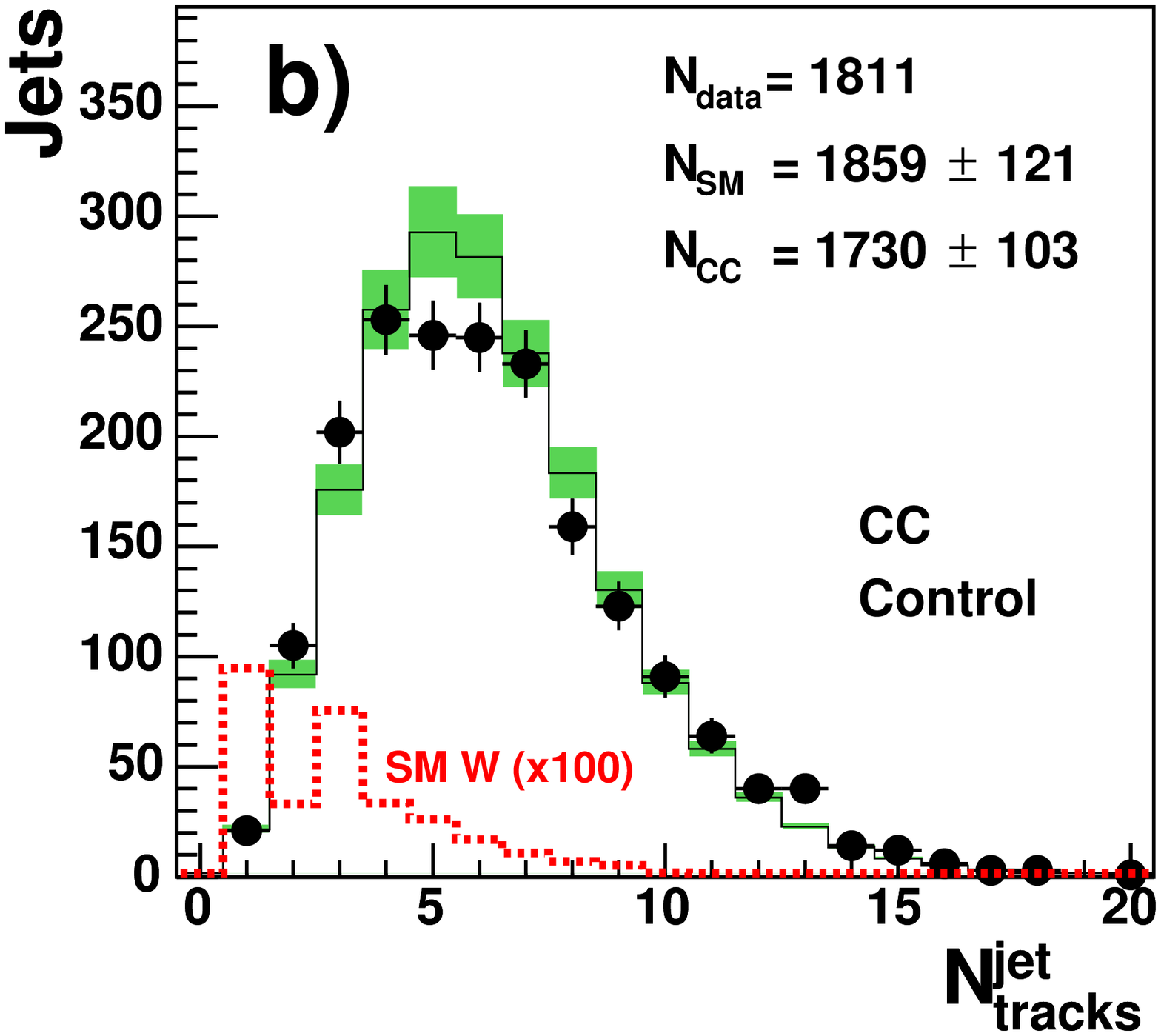,
                          bbllx=-2,bblly=1,bburx=577,bbury=574, angle=0, clip=, width=6.5cm}}}
\put(0,55){\mbox{\epsfig{file = 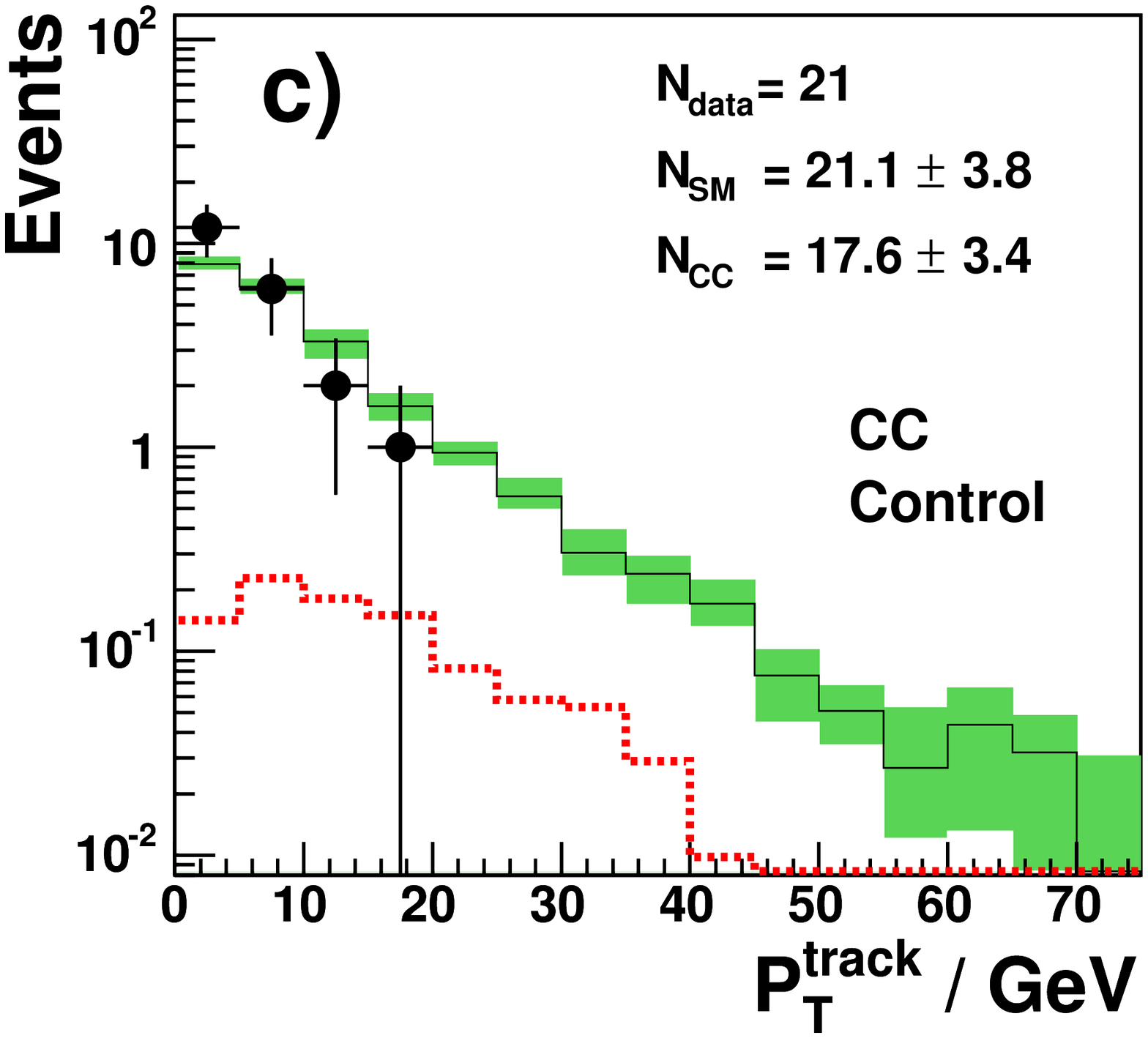,
                          bbllx=-2,bblly=1,bburx=577,bbury=574, angle=0, clip=, width=6.5cm}}}
\put(70,55){\mbox{\epsfig{file = 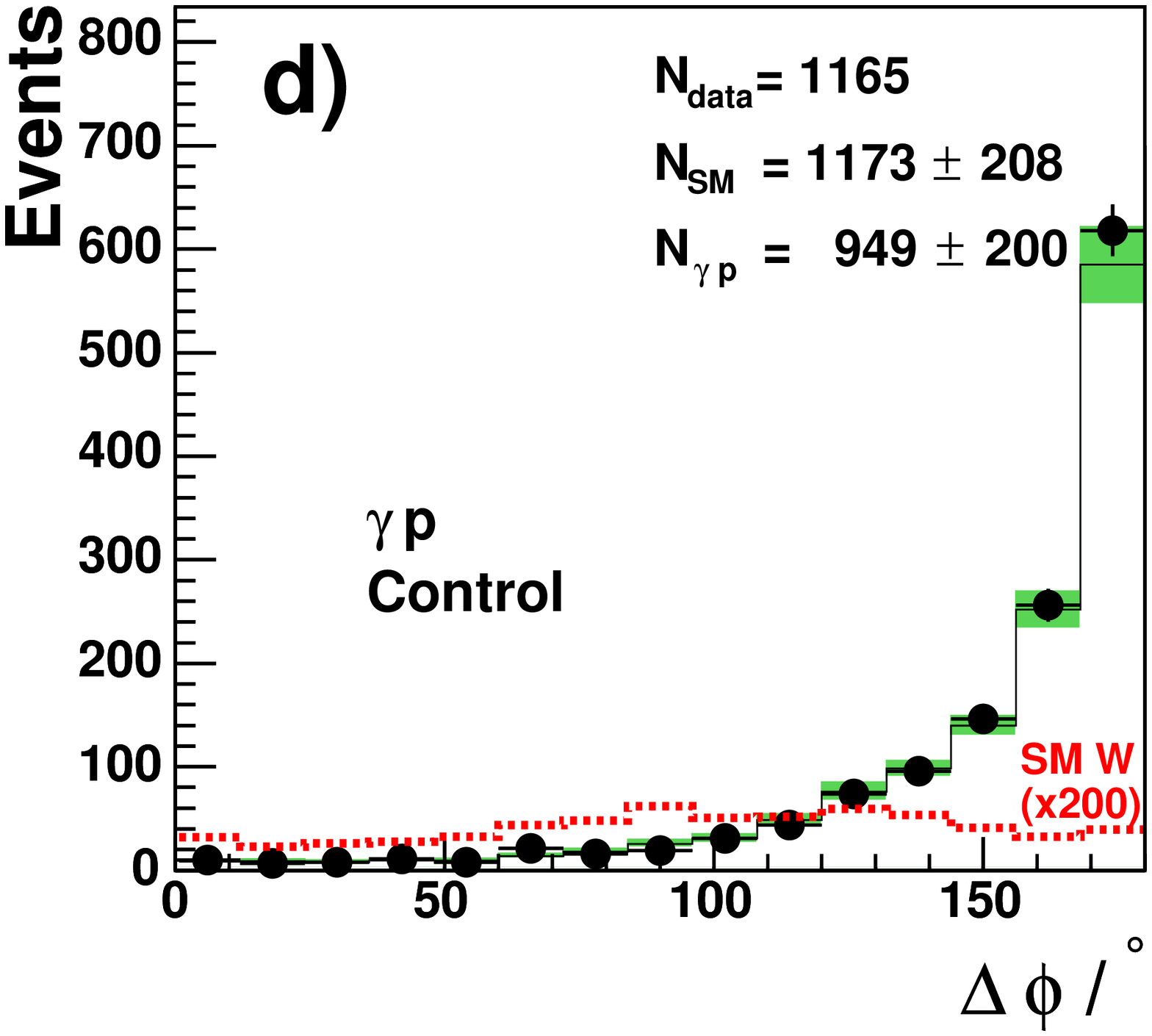,
                          bbllx=-2,bblly=1,bburx=577,bbury=574, angle=0, clip=, width=6.5cm}}}
\put(0,0){\mbox{\epsfig{file = 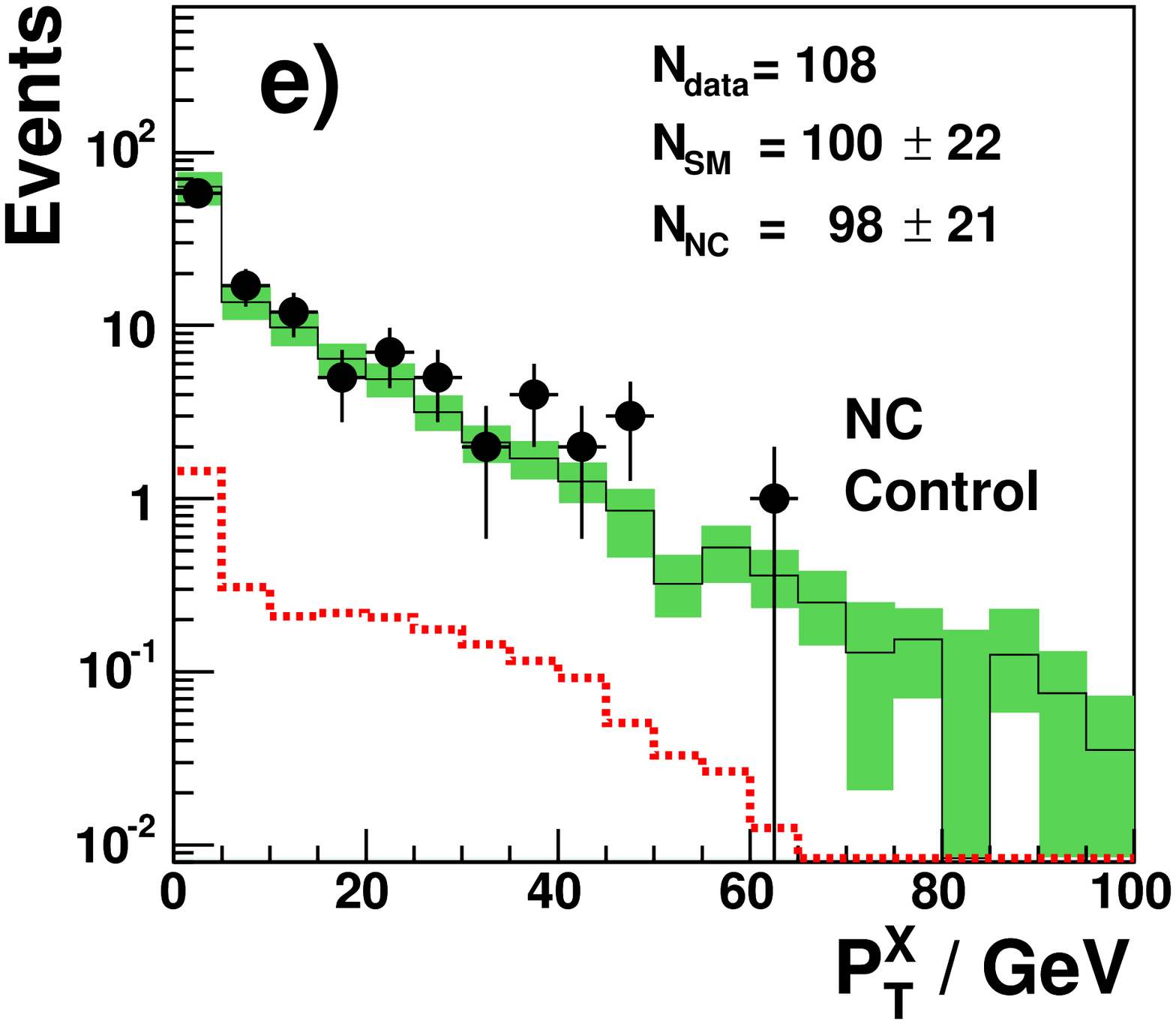,
                          bbllx=-2,bblly=1,bburx=577,bbury=574, angle=0, clip=, width=6.5cm}}}
\put(70,10){\mbox{\epsfig{file = 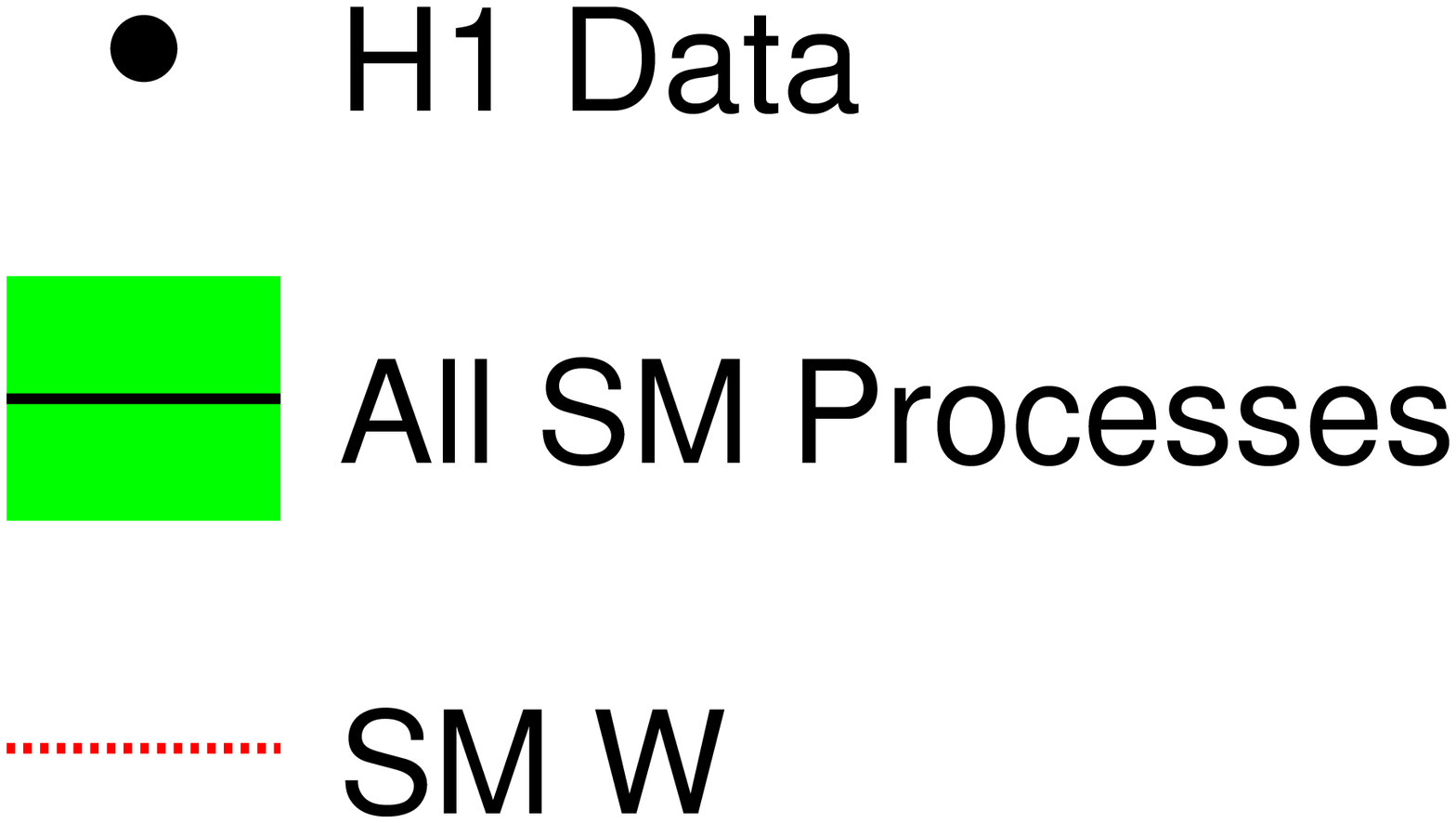,
                          bbllx=-2,bblly=1,bburx=577,bbury=574, angle=0, clip=, width=5.8cm}}}

\end{picture}
\end{center}
\caption{Control distributions in the background enriched samples defined for the  $\tau+P_T^\mathrm{miss}$ analysis:
          {\bf a)} the radius of isolated jets, {\bf b)} the number of charged particles in each isolated jet  and {\bf c)} the transverse momentum of the track associated to single track isolated jets in the CC DIS control sample; {\bf d)} the 
         acoplanarity angle between the  isolated jet and the remaining hadronic system X  in the $\gamma p$ control sample; {\bf e)} the transverse momentum of the hadronic system X excluding the isolated jet candidate  in the NC DIS control sample.
         In all distributions each event may enter several times, 
         depending on the number of isolated jets in the event. In each figure the data (points) are  compared to the SM expectation (solid histogram) shown with its uncertainty (shaded band). The signal contribution dominated by the SM $W$ production is also shown (dashed histogram). In  {\bf a)} and  {\bf b)} it is shown multiplied by a factor 100 and in {\bf d)} by 200. }
\label{backgr}
\end{figure}

As shown in figure~\ref{backgr}a--c, the distributions of $R_{jet}$, $N_{tracks}^{jet}$, and  $P_T^{track}$  in the CC DIS control sample are well described by the MC simulation, both in shape and normalisation.
The distribution of the acoplanarity angle $\Delta \phi$ in the control sample enriched in $\gamma p$ events, shown in figure~\ref{backgr}d, shows a clear peak towards 180$^{\circ}$, corresponding to back--to--back events.
In figure~\ref{backgr}e the distribution of the hadronic transverse momentum $P_{T}^{X}$ for the NC DIS control sample is shown.
The good agreement between data and MC simulation observed in all control samples confirms the good understanding of background rates and of the properties of the jets (shape and multiplicity) used in the tau identification procedure.

\renewcommand{\arraystretch}{1.5}
\begin{table}[hhh]
\begin{center}
\begin{tabular}{|l|l|l|l|}
\hline
\multicolumn{4}{|c|}{\bf \boldmath{$\tau+P_T^\mathrm{miss}$} Background Control Samples} \\ 
\hline
\multirow{2}{6mm}{\bf Control \\Sample} &
  & & \\
& \multicolumn{1}{|c|}{\bf CC DIS} & 
\multicolumn{1}{|c|}{\bf NC DIS} &
\multicolumn{1}{|c|}{\bf \boldmath{$\gamma p$ } }\\
\hline \hline
\multirow{4}{10mm}{Selection} & \multicolumn{3}{|c|}{ $P^{\rm miss}_T> 12$ GeV,   $\;\;$
 $P^{\rm calo}_T> 12$ GeV, $\;\;$ at least one isolated jet } \\ \cline{2-4} 
&  $E-P_z< 50$ GeV & $E-P_z> 35$ GeV & $ E-P_z< 50$ GeV \\
& $V_{ap}/V_{p} <0.5 $ & $P_T^e > 10$ GeV & $V_{ap}/V_{p}> 0.15$
 \\
& $V_{ap}/V_{p} <0.15 \;\mathrm{for}\; P_T^{\rm calo}<25$ GeV & $5^\circ<\theta^e<150^\circ$   & no electrons
 \\
\hline \hline
Data & $1811$ &  $108$ &  $1165$ \\

Total SM & $1858.6 \pm 120.7 $ &  $ 100.3 \pm 22.0 $ &  $1173.1 \pm 208.3$ \\
 & (93\% CC) & (98\% NC) & (80\% $\gamma P$) \\
 \hline

\end{tabular}
\caption{ Selection criteria for the $\tau+P_T^\mathrm{miss}$ background control samples 
         together with the number of selected events compared to the SM expectation. The dominant contribution to the SM prediction is indicated as a percentage in the last row of the table.}
\label{backgrcut}
\end{center}
\end{table}

\subsection{Systematic uncertainties}
\label{sec:syserr}
In this section, the systematic uncertainties associated with  the search for events containing  an isolated tau lepton and  large missing transverse momentum
are discussed.
The effect of those uncertainties on the expectations from SM W production and from background processes is determined by varying the experimental quantities by $\pm 1$ standard deviation.

\begin{itemize}
%%%%%%%%%%%%%%%%%%%%%%%%%%%%%%%%%%%%%%%%%%%%%%%%%%%%
\item {\bf Tau identification procedure}\\
The main experimental uncertainty on the signal and background expectations arises from the requirement of exactly one track within the jet and from the condition on the jet radius $R_{jet} < 0.12$. The energy of charged tracks is reconstructed with an uncertainty of $5\%$. The measurements of the polar and azimuthal angles are attributed a systematic error of 3~mrad and 1~mrad respectively. The uncertainty on the track reconstruction efficiency is $3\%$ and has a relatively large effect on the background expectation, due to migrations of hadronic jets with multiple charged particles into the single track category. 
The modelling of $R_{jet}$ is studied with high statistical precision using jets in a dedicated inclusive NC DIS sample.
%, selected by requiring one well identified electron detected in the LAr calorimeter.
The measured and simulated jet radius distributions are compared 
and an uncertainty depending on the polar angle of the $\tau$--jet is attributed to $R_{jet}$. 
The uncertainties associated with the tau identification are  $15\%$ for the expected $W$ signal and  $25\%$ for the background predictions.
%%%%%%%%%%%%%%%%%%%%%%%%%%%%%%%%%%%%%%%%%%%%%%%%%%%%
\item {\bf Hadronic final state reconstruction}\\
The hadronic energy measurement has a relative uncertainty of $4\%$. The polar angle measurement has an uncertainty varying from 3 mrad for jets reconstructed in the central region to 10 mrad for forward jets. The topological variable $V_{ap}/V_{p}$ is attributed  a relative uncertainty of $10\%$.
%%%%%%%%%%%%%%%%%%%%%%%%%%%%%%%%%%%%%%%%%%%%%%%%%%%%
\item {\bf Trigger} \\
The uncertainty on the $P_T^\mathrm{miss}$ trigger efficiency is $5\%$ deduced from a large statistics  NC DIS sample for which the trigger information is reconstructed offline, ignoring the signal from the electron~\cite{Adloff:2000qj}.
%%%%%%%%%%%%%%%%%%%%%%%%%%%%%%%%%%%%%%%%%%%%%%%%%%%%
\item {\bf Luminosity} \\
The luminosity of the analysed datasets is measured with an uncertainty of $1.5\%$.
%%%%%%%%%%%%%%%%%%%%%%%%%%%%%%%%%%%%%%%%%%%%%%%%%%%%
\item {\bf Theoretical errors on signal and background contributions}\\
The uncertainty on the $W$ production signal cross section is estimated to be $15\%$~\cite{SPIRA}.
The expectations from NC DIS and $\gamma p$ processes are each attributed an additional uncertainty of $20\%$,
a value which has been estimated from the control samples described in section~\ref{sec:backgstudy}. This uncertainty covers the sensitivity of the jet radius and multiplicity (used in the tau identification algorithm)  to the modelling of parton showers in NC DIS and $\gamma p$ MC samples.

\end{itemize}

The individual effects of the experimental uncertainties are combined in quadrature to give the total experimental systematic uncertainty. 
The total uncertainty on the SM W signal is $20\%$  and that on the SM background  is $34\%$.  For both signal and background processes, the total uncertainty is dominated   by the uncertainty arising from the tau identification procedure and the theoretical uncertainties.

\subsection{Results}

\par
\begin{figure}[hhh]
\setlength{\unitlength}{1mm}
\begin{center}
\begin{picture}(150,140)(0,0)
\put(-10,70){\mbox{\epsfig{file = 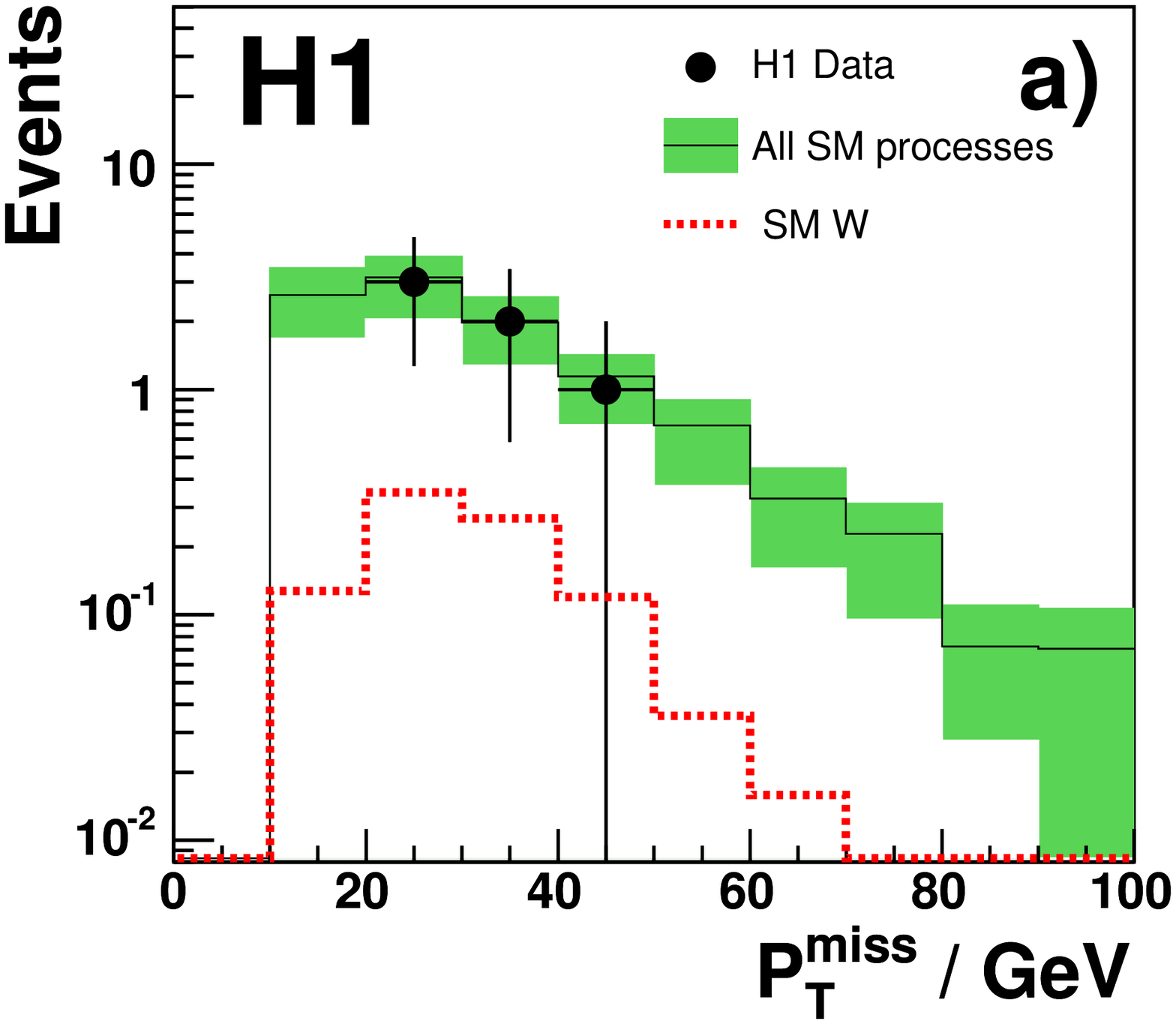,
                          bbllx=-2,bblly=1,bburx=577,bbury=574, angle=0, clip=, width=8.5cm}}}
\put(73,70){\mbox{\epsfig{file = 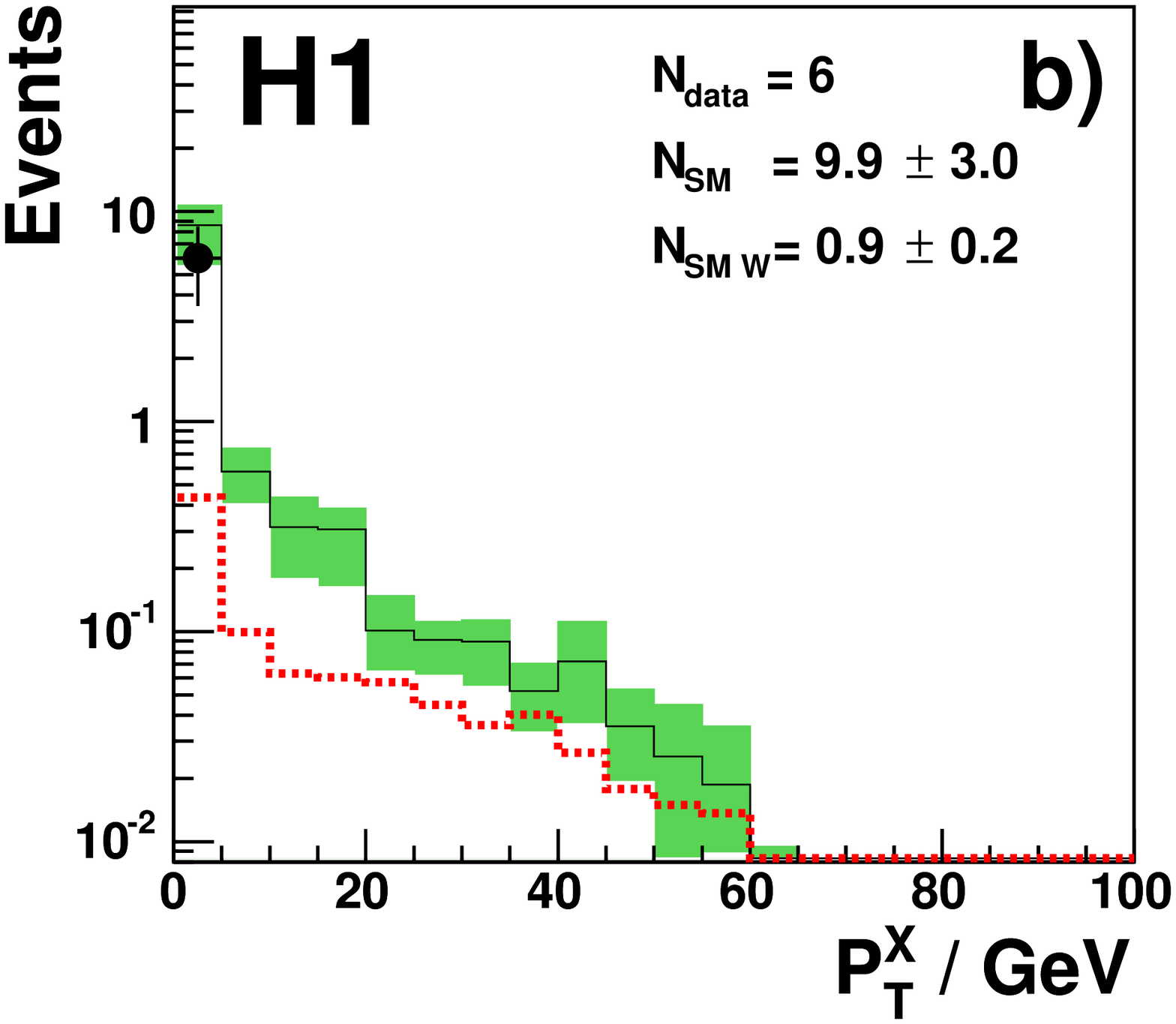,
                          bbllx=-2,bblly=1,bburx=577,bbury=574, angle=0, clip=, width=8.5cm}}}
\put(-10,0){\mbox{\epsfig{file = 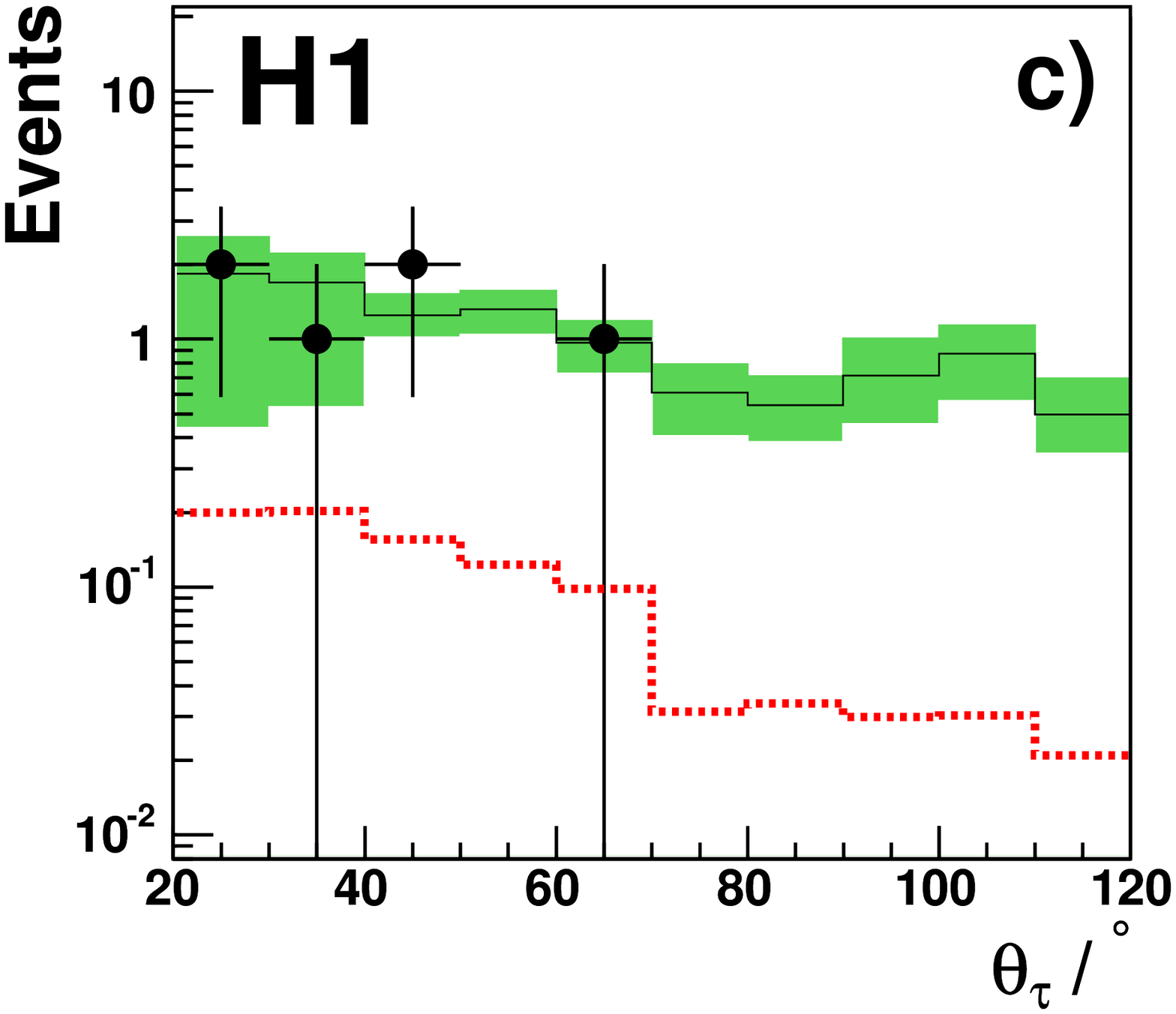,
                          bbllx=-2,bblly=1,bburx=577,bbury=574, angle=0, clip=, width=8.5cm}}}
\put(73,0){\mbox{\epsfig{file = 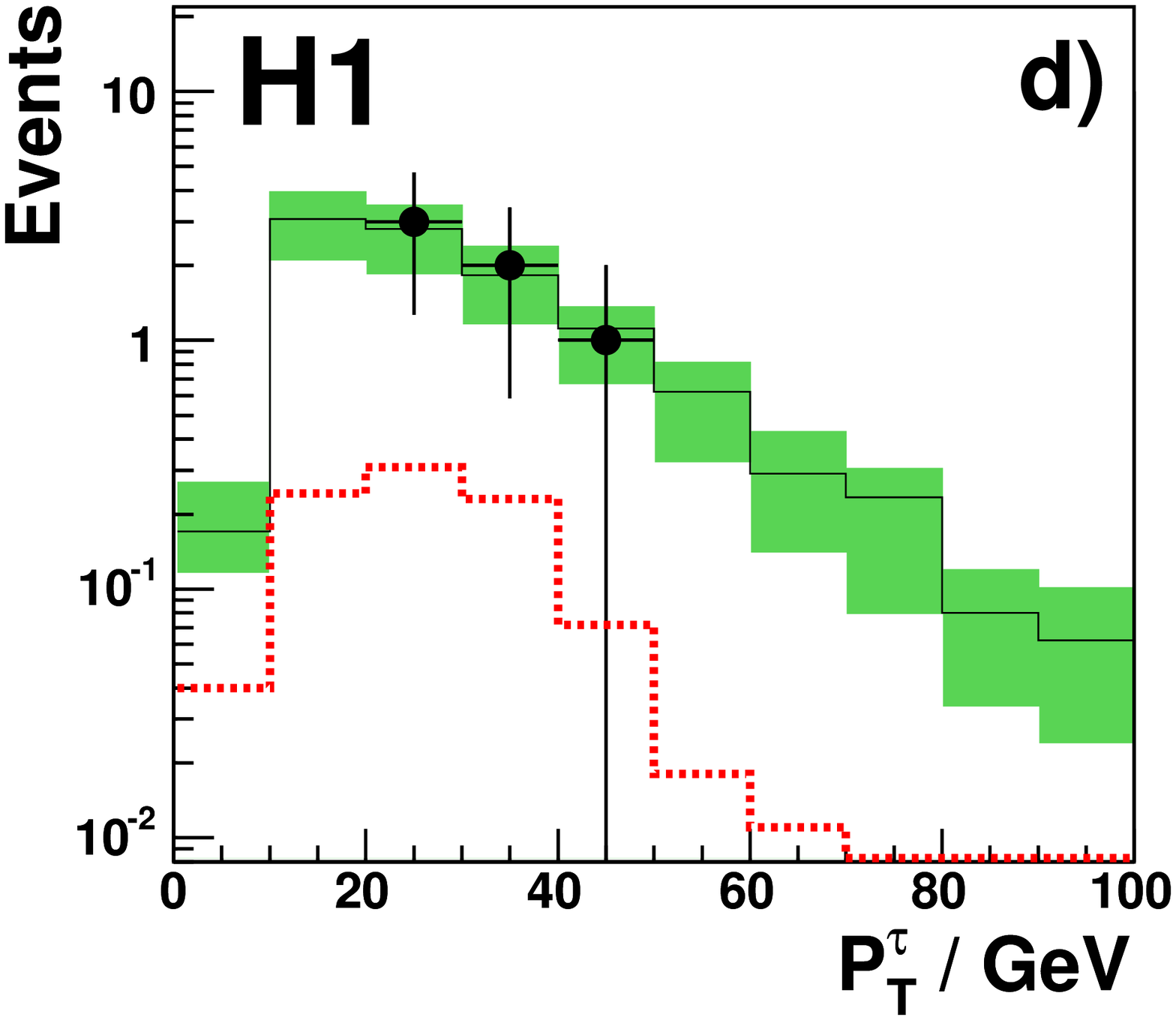,
                          bbllx=-2,bblly=1,bburx=577,bbury=574, angle=0, clip=, width=8.5cm}}}

\end{picture}
\end{center}
\caption{Distributions of 
{\bf a)} the total missing transverse momentum $P^\mathrm{miss}_T$,
{\bf b)} the hadronic transverse momentum $P_{T}^{X}$, 
{\bf c)} the polar angle $\theta_{\tau}$   and
{\bf d)} the transverse momentum $P_T^{\tau}$ of the $\tau$--jet candidate
         in the final sample of events containing an isolated tau lepton and large missing transverse momentum. In each figure the data (points) are  compared to the SM expectation (solid histogram) shown with its uncertainty (shaded band). The signal contribution dominated by the SM $W$ production is also shown (dashed histogram).}
\label{isotaufinal}
\end{figure}

In the final event sample $6$ events are observed in the data,
compared to a total SM expectation of $9.9$~$^{+ 2.5}_{- 3.6}$ events, of which $0.89^{+ 0.15}_{- 0.26}$ are expected from SM $W$ production.
The $P_T^{X}$ spectrum and other properties of the events in the final sample are shown in figure~\ref{isotaufinal}. 
Table~\ref{isotauresults} summarises the results.
The events observed in the data are concentrated in the region of very low $P_T^X$,
where the contribution from CC DIS background processes dominates the SM expectation.
In the region $P_T^X > 25$~GeV, where an excess of events containing isolated electrons or muons is observed~\cite{Andreev:2003pm}, no event is found for a SM prediction of $0.39 \pm 0.10$, of which $0.20\pm0.04$ are expected from SM $W$ production.

\par
In the absence of a signal, a model independent upper limit on the cross section for the production of events containing an isolated tau leptons and large missing transverse momentum is set
in the kinematic region: 
$5^{\circ}$~$<$~$\theta_{\tau}$~$<$~$140^{\circ}$, $P_{T}^{\tau} > 10$~GeV and $P^{\rm miss}_T > 12$~GeV.
The limit is calculated using a modified frequentist approach based on likelihood ratios~\cite{tlimit} and taking into account the systematic uncertainties discussed in section~\ref{sec:syserr}.
The acceptance for processes producing isolated tau leptons in events with large missing transverse momentum in the given kinematic region
 is estimated using the MC simulation for SM $W$ production,
implemented in the EPVEC generator. 
 An additional model uncertainty of $10\%$ is attributed to the acceptance. This uncertainty is estimated by comparing the acceptance predicted by EPVEC with that obtained using the generator ANOTOP, which simulates the anomalous production of single top quarks in $ep$ collisions at HERA~\cite{ANOTOP} and produces $W$ bosons with different
kinematic distributions\footnote{The $P_T^X$ distribution for top events is peaked at about $70$~GeV, while events from SM W production are concentrated at low  $P_T^X$ values. }.

\par
An upper limit of $\sigma < 0.85$~pb at $95\%$ confidence level is obtained for the production cross section of events containing an isolated tau lepton and large missing transverse momentum in the phase space defined above.
In the region $P^X_T > 25$~GeV,
the upper limit is \linebreak $\sigma(P_T^X>25~\mathrm{GeV}) < 0.31$~pb at $95\%$ confidence level.
These limits are higher than the cross sections measured in the electron and muon channels~\cite{Andreev:2003pm}. The present  measurement is therefore compatible with the previous measurement of events with an electron or muon and $P_T^\mathrm{miss}$, as expected if lepton universality is assumed.
\renewcommand{\arraystretch}{1.8}  
\begin{table}
\begin{flushleft}

\begin{tabular}{|c||c|c||c|c|}
\hline
{\boldmath{$\tau+P_T^\mathrm{miss}$}} {\bf Results} & H1 Data  & SM Expectation & SM Signal & Other SM Processes \\
\hline
Total        &   $6$  &  $9.9$ $^{+ 2.5}_{- 3.6}$   &  $0.89$ $^{+ 0.15}_{- 0.26}$  &  $9.0$ $^{+ 2.5}_{- 3.6}$ \\
\hline
$P^X_T > 25$~GeV   &   $0$  &  $0.39$ $^{+ 0.09}_{- 0.11}$    &  $0.20$ $^{+ 0.04}_{- 0.05}$  &  $0.19$ $^{+ 0.08}_{- 0.10}$ \\
\hline
$P^X_T > 40$~GeV   &   $0$  &  $0.16$ $^{+ 0.07}_{- 0.06}$    &  $0.08$ $^{+ 0.02}_{- 0.02}$  &  $0.08$ $^{+ 0.07}_{- 0.06}$ \\
\hline

\end{tabular}
\end{flushleft}
\caption{Number of events observed in the data compared to the SM expectation for signal (W decay into $\tau$) and background processes
         in the final $\tau + P^\mathrm{miss}_T$ sample.}
\label{isotauresults}
\end{table}
%\newpage

%======================================
\section{Summary} 
%======================================

In this paper, the production of  tau leptons  in $ep$ collisions at HERA is investigated in events containing a $\tau^{+} \tau^{-}$ pair and events containing an isolated tau lepton and large missing transverse momentum.

\par
The production cross section of  $\tau^{+} \tau^{-}$ pairs is measured in elastic processes, in a combination of
leptonic, semi--leptonic and hadronic decay modes of the two tau leptons. 
In a data sample corresponding to an integrated luminosity of $106$~pb$^{-1}$,
$30$ events are observed, in agreement with a Standard Model expectation of $27.1 \pm 4.1$ events,
of which $16.0 \pm 3.4$ are expected from $\gamma \gamma \rightarrow \tau^{+} \tau^{-}$ signal processes.
This is the first observation of tau pair production in $ep$ collisions.
\par
A search for  the production of isolated tau leptons in events with large missing transverse momentum is  performed in a data sample corresponding to an integrated luminosity of $115$~pb$^{-1}$. The selection yields 
$6$ candidate events, compatible with a Standard Model expectation of $9.9$~$^{+ 2.5}_{- 3.6}$ events.
No event is observed in the region $P_T^X>25~\gev$, where $0.39 \pm 0.10$ events are expected, including $0.20\pm0.05$ events from $W\rightarrow \tau \nu_\tau$ decays.
An upper limit on the production cross section of $\sigma(P_T^X>25~\mathrm{GeV})$~$<$~$0.31$~pb at $95\%$~confidence level is set.

%+++++++++++++++++++++++++++++++++++++++++++++++++++++++++++++++++
\section*{Acknowledgements} 
%++++++++++++++++++++++++++++++++++++++++++++++++++++++++++++++++++++

We are grateful to the HERA machine group whose outstanding efforts have made this experiment
possible. We thank the engineers and technicians for their work in constructing and
maintaining the H1 detector, our funding agencies for financial support, the DESY technical
staff for continual assistance and the DESY directorate for support and for the hospitality which
they extend to the non--DESY members of the collaboration.

\end{document}